\documentclass[twocolumn]{aastex62}
\pdfoutput=1 
\usepackage{amsmath,amssymb,amstext}
\usepackage{gensymb}

\submitjournal{ApJ}

\shorttitle{Simulating Outflows from the LMC}
\shortauthors{Bustard et al.}

\begin{document}

\title{Cosmic Ray Driven Outflows from the Large Magellanic Cloud: Contributions to the LMC Filament}

\correspondingauthor{Chad Bustard}
\email{bustard@wisc.edu}
\author{Chad Bustard}
\affil{Department of Physics, University of Wisconsin-Madison, 1150 University Avenue, Madison, WI 53706}

\author{Ellen G. Zweibel}
\affil{Department of Physics, University of Wisconsin-Madison, 1150 University Avenue, Madison, WI 53706}
\affiliation{Department of Astronomy, University of Wisconsin - Madison, 475 North Charter Street, Madison, WI 53706, USA}

\author{Elena D'Onghia}
\affiliation{Department of Astronomy, University of Wisconsin - Madison, 475 North Charter Street, Madison, WI 53706, USA}

\author{J.S. Gallagher III}
\affiliation{Department of Astronomy, University of Wisconsin - Madison, 475 North Charter Street, Madison, WI 53706, USA}

\author{Ryan Farber}
\affiliation{Department of Astronomy, University of Michigan, 1085 South University, Ann Arbor, MI 48109}

\begin{abstract}
In this paper, we build from previous work (Bustard et al. 2018) and present simulations of recent (within the past Gyr), magnetized, cosmic ray driven outflows from the Large Magellanic Cloud (LMC), including our first attempts to explicitly use the derived star formation history of the LMC to seed outflow generation. We run a parameter set of simulations for different LMC gas masses and cosmic ray transport treatments, and we make preliminary comparisons to published outflow flux estimates, neutral and ionized hydrogen observations, and Faraday rotation measure maps. We additionally report on the gas mass that becomes unbound from the LMC disk and swept by ram pressure into the Trailing Magellanic Stream. We find that, even for our largest outburst, the mass contribution to the Stream is still quite small, as much of the outflow-turned-halo gas is shielded on the LMC’s far-side due to the LMC’s primarily face-on infall through the Milky Way halo over the past Gyr. On the LMC's near-side, past outflows have fought an uphill battle against ram pressure, with near-side halo mass being at least a factor of a few smaller than the far-side. Absorption line studies probing only the LMC foreground, then, may be severely underestimating the total mass of the LMC halo formed by outflows. 

\end{abstract}


\section{Introduction}

The Large and Small Magellanic Clouds (LMC and SMC, respectively), two of the Milky Way's nearby dwarf satellite galaxies, present a unique laboratory for studying gas dynamics and galaxy evolution. While falling into the Milky Way halo, the Clouds orbit around each other, triggering bursts of star formation and also tearing each other apart as their gravitational forces loosen and strip material. Combined with the constant headwind they experience during their infall, their galactic dance flings gas behind the Clouds, contributing to the Trailing Magellanic Stream (see \cite{ElenaReview2016} for a recent review). This massive gaseous tail extending hundreds of kpc behind the Clouds is an intriguing component of our Milky Way circumgalactic medium (CGM), as it may someday fall onto the Milky Way disk and enhance our Galactic ecosystem by providing more fuel to form stars. Fortunately, because of our birds-eye view, the Magellanic System gives us an incredible window into how galaxies expel and feed on gas; both the large-scale gas cycles in and between galaxies, as well as the small-scale, internal processes that drive gas flows. 

From recent proper motion measurements \citep{Kallivayalil2006, 2013ApJ...764..161K}, we can infer that the LMC and SMC are on their first or second infall into the Milky Way halo \citep{Besla2007}. In the prevailing first infall scenario, dwarf-dwarf galaxy interaction models can recreate much of the observed morphology of the Stream, primarily due to the larger LMC tidally stripping the SMC \citep{Besla2010, Besla2012}. This is supported by observations tying the chemical enrichment of the Stream predominantly back to the SMC \citep{Fox2013}; however, some unsolved puzzles still remain. The overall mass of the Stream, taking into account the significant amount of ionized hydrogen first detected in H$\alpha$ by \cite{Weiner1996HAlpha} and now in greater detail by the Wisconsin H$\alpha$ mapper (WHAM) \citep{Barger2017}, exceeds that produced by tidal stripping models; even when the LMC and SMC are simulated with more gas-rich disks, the resulting mass underestimates the observed mass ($\approx 2 \times 10^{9} M_{\odot}$; \cite{Fox2014}) by a factor of four \citep{Pardy2018}. Additionally, the Stream exhibits two bifurcated tails offset both kinematically \citep{Nidever2008TheArm} and chemically \citep{Richter2013}, with one leading back to the LMC that is not as easily reproducable by purely tidal interaction models. 
Given these puzzles, it is prudent to consider additional gas physics beyond tidal models, in hopes that one can explain observed properties of the Magellanic System and, in turn, learn about the role that these processes play in galaxy evolution more broadly. 

To this end, we focus our attention on two pieces of the complex Magellanic environment: 1) Supernova-driven outflows from the LMC and 2) The LMC filament, an extended trail of gas, offset in both velocity and abundance from the rest of the Stream. Interestingly, the two may be related, as the filament leads back to an active star forming region of the LMC, the Southeast H I Overdensity \citep{Nidever2008TheArm}, that may be energizing outflows from the disk. Indeed, the LMC observed in H I is dotted with holes coincident with hot x-ray emitting shells \citep{Kim1999HICloud, Kim2003AS}, likely the remnants of past supernovae and gas ejection episodes. Absorption line studies \citep{Howk2002, Lehner2007} show evidence for the culmination of this recent activity: a multiphase, large-scale outflow, with recent evidence for its existence on both sides of the disk \citep{Barger2016}. Best estimates of the outflow mass flux and velocity are $0.4 M_{\odot}/yr$ and $\approx 100$ km/s from \cite{Barger2016}. Motivated by these observations, we set out to simulate these outflows, their mass contribution to the broader LMC CGM, and, as hypothesized by \cite{Nidever2008TheArm}, whether they can comprise the LMC filament.

In \cite{Bustard2018} (hereafter referred to as B2018), we simulated supernova-driven outflows from the LMC and their interaction with an edge-on ram pressure, the force the LMC feels as it falls into the Milky Way halo. Using the FLASH magnetohydrodynamics (MHD) code \citep{FLASHRef}, we set up a wind-tunnel like simulation and showed that, in the Milky Way halo environment, even small “fountain” flows with velocities and mass fluxes comparable to observations \citep{Barger2016} can be expelled by ram pressure instead of falling back onto the disk. This gas then trails behind the LMC, forming a filament when viewed face-on. 

Given the positive explanatory power of these results, we now improve our simulations to more accurately model LMC outflows, their composition, and their interaction with a time-varying 3D ram pressure headwind. One important addition is cosmic ray wind driving, which notably affects the velocity, mass-loading, and fate of our LMC outflows. Cosmic rays, the most energetic, non-thermal particles in the universe, have been shown to be extremely effective in driving winds \citep{1991A&A...245...79B, 2008ApJ...674..258E,2014MNRAS.437.3312S, 2016ApJ...816L..19G, Ruszkowski2017}, and their unique signatures on the multiphase structure of the outflow and surrounding CGM are coming into focus \citep{Salem2016, 2018MNRAS.479.3042G, 2018ApJ...868..108B, Ji2019CRs}. 

We will analyze cosmic ray driven outflows from the LMC using two different modes of cosmic ray transport: \emph{advection}, whereby the cosmic ray fluid is locked to the gas, and \emph{streaming} with additional collisional loss terms as well. We will discuss preliminary differences between outflows given each transport model, but save a deeper analysis of the streaming simulations for future work (Bustard et al., in prep). 

We also use this opportunity to outline some of our first attempts to explicitly include the derived star formation rate (SFR) history of the LMC \citep{Harris2009THECLOUD} into our simulations. Compared to B2018, this is a large step for assessing the past and present day properties of LMC outflows and their possible formation of the LMC filament. This SFR, along with recent estimates of LMC wind mass flux and velocity \cite{Barger2016}, neutral and ionized hydrogen column density maps, and gamma-ray observations, may give us a lever to determine the LMC gas mass within the past Gyr, as well as the nature of cosmic ray transport in the LMC environment.  

Our paper is outlined as follows: We first describe in Section \ref{sec:sectionSetup} our LMC initial conditions, the magnetic field configuration, our choice of Milky Way halo parameters, the LMC orbit parameters that determine the ram pressure headwind velocity and angle, and the star formation history of the LMC. In Section \ref{sec:SFRHistory}, we describe how we seed outflows from the derived star formation history of the LMC. In Section \ref{methods}, we outline the MHD equations we solve numerically for both the thermal gas and cosmic ray fluids, as well as our implementation of radiative cooling and gas equation of state. In Section \ref{sec:results}, we show our results, first isolating feedback and ram pressure separately and then putting them together. We then discuss our results in Section \ref{conclusions}, both in the context of the Magellanic System and also more broadly.



\section{Simulation Setup}
\label{sec:sectionSetup}

\subsection{LMC Disk Setup}
\label{sec:initialsetup}
We generally follow the simulation setup of \citet{Salem2015RAMMEDIUM}. Our initial gas distribution follows the form given in \citet{Tonnesen2009GASMEDIUM},
\begin{equation}
\label{gasDensityEquation}
\rho (R,z) = \rm \frac{M_{gas}}{2\pi a_{gas}^{2} b_{gas}} 0.5^{2} \rm sech\Big(\frac{R}{a_{gas}}\Big) \rm sech\Big(\frac{|z|}{b_{gas}}\Big) 
\end{equation} 
where $a_{gas}$ and $b_{gas}$ are the radial scale length and vertical scale height of the disk. This density is smoothly cutoff between a radius of $R_{\rm cut} = 10$ kpc and 13 kpc. This density distribution is then put in dynamical equilibrium with fixed stellar and dark matter gravitational potentials. Note that, as in \cite{Bustard2018}, we do not include the self-gravity of the gas. We model the stellar potential of the disk as a static Plummer-Kuzmin disk \citep{1975PASJ...27..533M} potential:
\begin{equation}
\label{stellarMassEquation}
\Phi(R,z) = GM_{\star} \left[R^{2} + \left(a_{\star} + \sqrt{b_{\star}^{2} + z^{2}} \right)^{2} \right]^{-1/2}
\end{equation}
As in \citet{Salem2015RAMMEDIUM}, we use a total stellar mass of $3 \times 10^{9} M_{\odot}$, a radial scale length of $a_{\rm gas} = 1.7$ kpc and a vertical scale height of $b_{\rm gas} = 0.34$ kpc. We additionally include an NFW halo potential \citep{NFW1997}:

\begin{equation}
    d \Phi/dr = -\frac{GM(< r) r}{r^{3}}
\end{equation}
\begin{equation}
    M(< r) = M_{200}\big[\frac{\rm ln(1+x)-\frac{x}{1+x}}{\rm ln(1+c)-\frac{c}{1+c}}]
\end{equation}
or written another way, 
\begin{equation}
    M(< r) = 4 \pi \rho_{0} R_{s}^{3} [ \rm ln(1+x)-\frac{x}{1+x}]
\end{equation}
Here, $R_{200} = c R_{s}$, $x = rc/R_{200} = r/R_{s}$. Our parameter choices for $\rho_{0} = 3.4 \times 10^{-24} \rm g/cm^{3}$ and $R_{s} = 3$ kpc are converted to $M_{200}$ and $R_{200}$ using a concentration factor $c = 10$ and shown in Table \ref{propsTable}, along with other initial galaxy parameters. This NFW halo choice matches well with the potential used by \citet{Salem2015RAMMEDIUM}.

For this paper, we model two different LMC gas masses: $M_{\rm gas} = 5 \times 10^{8} M_{\odot}$ (low gas mass), which matches the initial condition of \cite{Salem2015RAMMEDIUM} and the present-day neutral hydrogen mass of the LMC within 4 kpc \citep{1998ApJ...503..674K}, and $M_{\rm gas} = 10^{9} M_{\odot}$ (high gas mass), which better accounts for the significant contribution of ionized species to the total gas mass (e.g. \cite{Brianna2019AAS}). This higher gas mass is also motivated by recent LMC-SMC interaction models, which using more extended, gas-rich disks, can better recreate the present-day Magellanic Stream gas distribution \citep{Pardy2018}.


\begin{table}[htb]
\begin{tabular}{ c  c }
\hline
\hline
Stars  \\
\citep{1975PASJ...27..533M} \\
$M_{\star}$  & $3 \times 10^{9} M_{\odot}$  \\
$a_{\star}$  & 1.7 kpc \\
$b_{\star}$  & 0.34 kpc  \\
\hline
Gas  \\
\citep{Tonnesen2009GASMEDIUM} \\
$M_{\rm gas}$ & $5 \times 10^{8} M_{\odot}$ (low gas mass) \\
$M_{\rm gas}$ & $1.0 \times 10^{9} M_{\odot}$ (high gas mass) \\
$a_{\rm gas}$ & 1.7 kpc \\
$b_{\rm gas}$ & 0.34 kpc  \\
\hline
Dark Matter Halo \\
\citep{1996ApJ...462..563N} \\
$M_{200}$ & $2.54 \times 10^{10} M_{\odot}$ \\
$R_{200}$ & 30 kpc  \\
\hline
\end{tabular}
\caption{Table of galaxy stellar, gas, and NFW halo parameters. $M_{\star}$ and $M_{\rm gas}$ are the stellar mass and gas mass parameters used in Equations \ref{gasDensityEquation} and \ref{stellarMassEquation}, respectively. The low gas mass LMC follows \citet{Salem2015RAMMEDIUM}, and the high gas mass LMC accounts for the significant presence of ionized gas in the LMC and its halo. The resulting \emph{total} gas masses of the disks are $\approx 7.5 \times 10^{8} M_{\odot}$ and $\approx 1.5 \times 10^{9} M_{\odot}$. $a_{\star}$ and $a_{\rm gas}$ are the stellar and gas scale lengths, which are assumed here to be equal. $b_{\star}$ and $b_{\rm gas}$ are the stellar and gas scale heights, which are assumed to be one-fifth of the scale lengths. The dark matter halo parameters are almost equivalent to that used by \citet{Salem2015RAMMEDIUM}, but we assume an NFW potential instead of a Burkert potential \citep{burkert1995}.}
\label{propsTable}
\end{table}


The pressure profile is chosen so that the pressure gradient balances the imposed gravitational force in the vertical direction. Our gas is then given a rotational velocity peaking near 80 km/s, which matches well with the observed rotation curve of the LMC \citep{rotationCurveOlsen2011, rotationCurveVanderMarel2014}, to counteract the remaining difference between the radial gravitational force and the radial pressure gradient. We note that this configuration is stable in the absence of ram pressure, radiative cooling, and wind launching. Turning on cooling and heating results in a fast temperature drop in the central region, where the initial temperature was $\approx 10^{5}$ K, and a fast heating of the colder, outer regions. The decrease in pressure near the galaxy's center results in a vertical collapse, while the sudden heating to $\approx 10^{4}$ K in the outer regions puffs up the disk at large radii. Feedback in the central disk is then responsible for pressurizing the ISM and increasing the scale height back towards its initial value.

Note that we do not include a pre-existing LMC CGM; the medium surrounding our LMC disk takes on the average density of the Milky Way halo, a $\beta$-profile function of galactocentric radius \cite{Salem2015RAMMEDIUM}. However, both observations and cosmological simulations suggest that LMC-mass dwarf galaxies will harbor a CGM with baryonic mass comparable to or greater than that within the galaxy itself \citep{wakker1998, Tumlinson2017, Christensen2018, Hafen2019}. Our simulations isolate the contribution of LMC outflows to such a CGM, but it is prudent to keep in mind that outflow propagation will be affected by the pre-existing CGM, which may have a density exceeding the Milky Way halo density for much of the past Gyr. We expand on these topics in Section \ref{conclusions}.

\subsection{Magnetic Field Configuration}
\label{sec:BField}
Magnetic fields are an integral part of our simulations, as they mediate the propagation of cosmic rays and, by themselves, represent an energetically significant component of the galaxy. One of the main goals of this work is to study the magnetization of the Trailing Stream and intergalactic medium / Milky Way halo due to the stretching of magnetic fields flux-frozen to stripped and outflowing gas from the LMC. Past simulations show the multiple effects of magnetic fields on stripping. Magnetic draping, when surrounding field lines drape around the stripped galaxy, can naturally lead to a bifurcated structure in the trailing stripped gas \citep{RuszkowskiMagneticDraping}, which holds a bit of promise for interpreting the twisting, filamentary Trailing Magellanic Stream. 

Magnetization of the disk, itself, leads to two countervailing forces: a restoring magnetic tension force that can help bind gas to the galaxy, but also magnetic buoyancy that can drive a vertical flux out of the potential well. Meanwhile, magnetization of the gas that \emph{does} get stripped may suppress mixing with halo gas and keep the filament more intact \citep{2014ApJ...795..148T, Berlok2019}. For this work, we choose to magnetize only the LMC disk and leave the effects of the Milky Way halo magnetic field to future work.




The magnetic field is initialized as the divergence-free TOR (toroidal) configuration from \citet{2014ApJ...795..148T} with a different vertical dependence so that the magnetic field drops off as $\rm sech(z/b_{gas})$ as the gas density does. The magnetic field is defined within the galaxy (for grid cells satisfying $r < R_{cut}$ and $|z| < 5b_{gas}$) as
\begin{equation}
\begin{split}
    & B_{z} = 0 \\
    & B_{x} = \rm a_{zf} e^{-6R_{cyl}/R_{cut}} y/R_{cyl} \times \rm sin(2.5R_{cyl}/R_{cut}) \\
    & B_{y} = \rm a_{zf} e^{-6R_{cyl}/R_{cut}} x/R_{cyl} \times \rm sin(2.5R_{cyl}/R_{cut})
    \end{split}
\end{equation}
where $R_{cyl} = \sqrt{x^{2} + y^{2}}$ is the cylindrical radius, and $a_{zf} = \rm a_{0}sech(z/b_{gas})$. This magnetic field configuration was chosen by \citet{2014ApJ...795..148T} to peak a few kpc from the galaxy center and then fall off gradually with radius until the cutoff. Near galaxy center, the magnetic field is purposely weak where the velocity field is changing very rapidly and could cause numerical issues. 

Setting $a_{0}$ determines our magnetic field strength in the galaxy midplane. For this work, we fiducially choose $a_{0} = 3.76 \times 10^{-6}$, which gives a peak field strength of around $4 \mu G$. This strength is motivated by Faraday rotation measure (RM) studies of the LMC \citep{2005Sci...307.1610G, 2012ApJ...759...25M}, which suggest an \emph{ordered} field strength closer to 1 $\mu G$, but this is sub-dominant compared to the random magnetic field (inferred strength of $\approx 3-4 \mu G$) not probed by Faraday RM and also unresolved in our simulations. A $4 \mu G$ maximum field more likely matches the total RMS magnetic field in the LMC disk, which is the important input for our simulations since the \emph{total} magnetic field strength dictates magnetic pressure support and the local Alfv\'{e}n speed, i.e. the cosmic ray streaming speed. Compression during disk collapse and in supernova remnant shells additionally induces local fluctuations to higher magnetic field strengths.


Outside the galaxy (for grid cells \emph{not} satisfying $r < R_{cut}$ and $|z| < 5b_{gas}$), 
\begin{equation}
    \begin{split}
        & B_{x} = 10^{-15} \rm G \\
        & B_{y} = B_{z} = 0
    \end{split}
\end{equation}
The choice to have the entire halo magnetic field in the x-direction should have no consequence on the results since the $10^{-15}$ G field is negligible compared to the field within the galaxy. The halo field is not set to exactly zero to avoid complications with the implementation of cosmic ray streaming in the FLASH cosmic ray module.

\subsection{Milky Way Halo Model}
\label{sec:setupMWHalo}
With their setup, and with an informed density cutoff of $0.03 \rm cm^{-3}$ below which all hydrogen is considered to be ionized instead of neutral, \citet{Salem2015RAMMEDIUM} found a good fit between their initial gas profile and the observed neutral hydrogen column density of the present-day LMC \citep{Kim1999HICloud}. Along the leading edge and towards the Trailing Stream, however, the observed HI column drops off and flattens, respectively, going out radially. Along the leading edge, this is a signature of ram pressure stripping that they reproduce in their simulations using the LMC inclination and orbits from a tidal interaction model \citep{Besla2012}. By comparing their ram pressure stripping simulations to the observed compression along this leading edge, they constrain the $\beta$- profile \citep{1998ApJ...497..555M} for the diffuse, ionized component of the Milky Way halo density:
\begin{equation}
    n(r) = n_{0} \Big[1 + \Big(\frac{r}{r_{c}}\Big)^{2}\Big]^{-3 \beta / 2}
\end{equation}
where $n_{0} = 0.46 \rm cm^{-3}$, $\beta = 0.559$, and $r_{c} = 0.35$ kpc are the best-fit parameters \citep{Salem2015RAMMEDIUM}. These parameters match well with observationally determined halo profiles (e.g. \cite{2019arXiv190909169F} and see Fig. 3 of B2018); therefore, we choose these same parameters for our Milky Way halo gas, and we model the LMC's ram pressure headwind by following \cite{Salem2015RAMMEDIUM} exactly.

\subsection{LMC Orbit and Ram Pressure Headwind}
\label{headwind}

\begin{figure}
    \centering
    \includegraphics[width=0.4\textwidth]{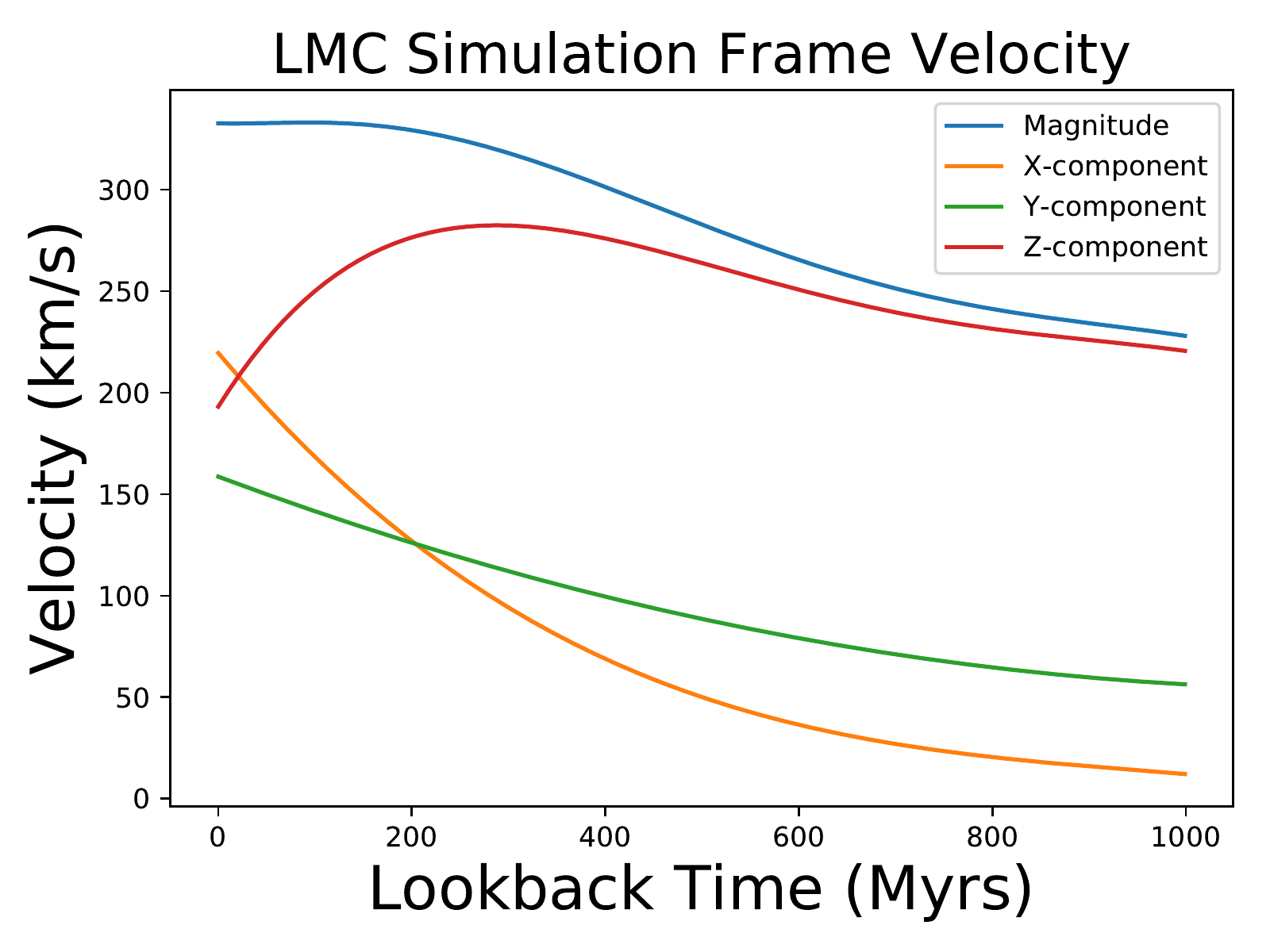}
    \includegraphics[width=0.4\textwidth]{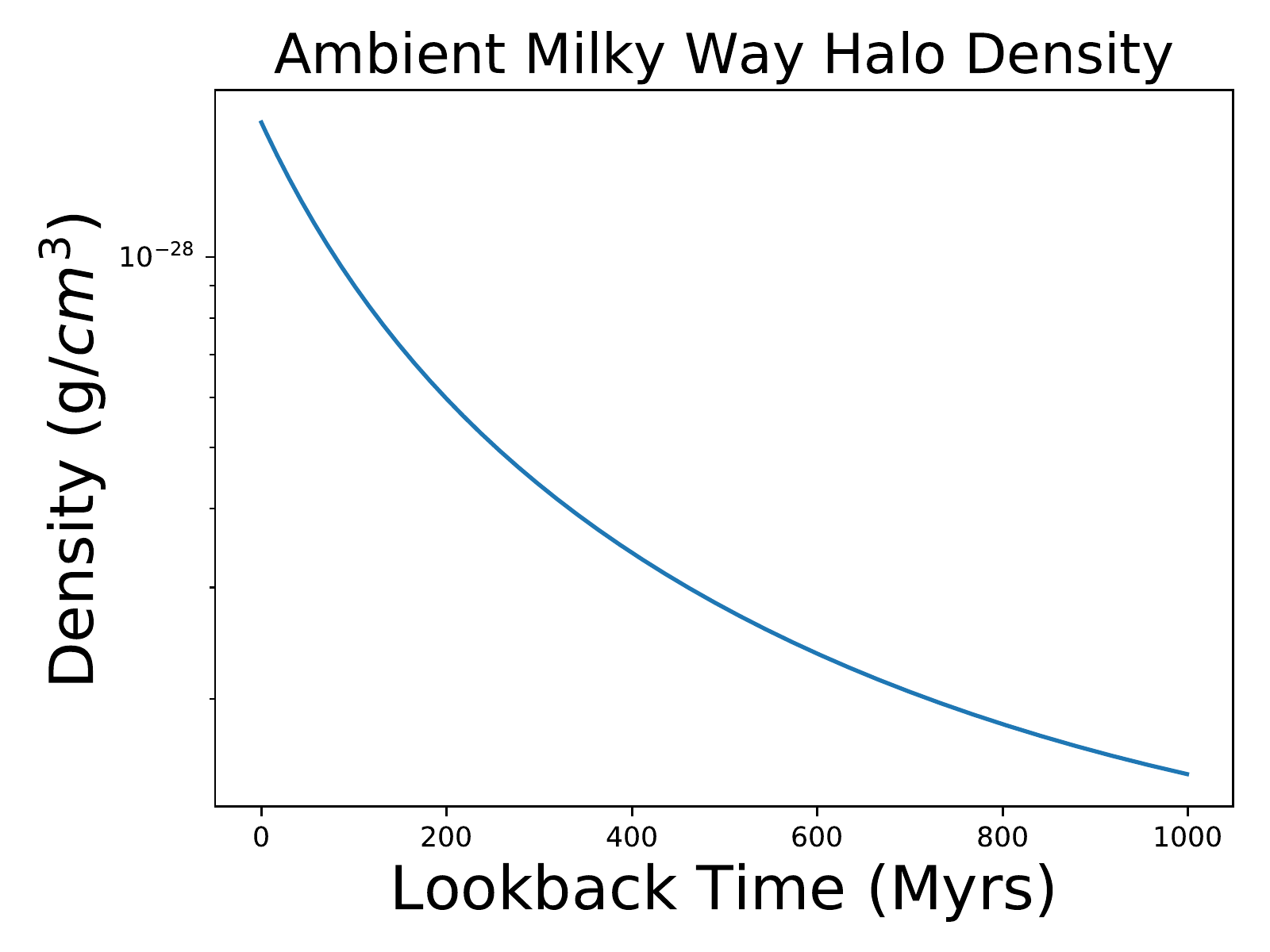}
    \caption{Ram pressure velocity components and density as a function of lookback time (t = 0.0 is present-day). Until the last few hundred Myrs, the LMC infall was primarily face-on ($\hat{z}$-direction), with the ambient density increasing from $\approx 10^{-29}$ to $10^{-28} \rm g cm^{-3}$ throughout the last Gyr assuming the LMC orbit from \cite{Besla2012} and the Milky Way halo density inferred from \cite{Salem2015RAMMEDIUM}.}
    \label{fig:RPHeadwind}
\end{figure}
To model the LMC's infall into the Milky Way halo, we sit in the frame of the LMC and turn on a wind-tunnel boundary condition from the box edges. Following \citet{Salem2015RAMMEDIUM}, we combine the density profile (above) with the orbital velocity and inclination of the LMC \citep{Besla2012} to get the in-flowing density and velocity vector as a function of time. Unlike in B2018, where we only considered the LMC to be infalling edge-on, we now include the time-dependent tilt of the LMC. Because the velocity vector now changes direction over time, it is wise to make a frame transformation such that the inflow only comes from 3 box edges \citep{Salem2015RAMMEDIUM}. This limits the propagation of inflows from opposite box edges, which could otherwise unrealistically overlap and form shocks throughout the simulation domain. This new transformed frame is the ``simulation frame" defined in \citet{Salem2015RAMMEDIUM}. We refer the reader to Table 3 of \citet{Salem2015RAMMEDIUM}, which outlines the transformations between reference frames that we follow in this work, as well as our Appendix that gives a brief overview of these frame transformations.

We see from Figure \ref{fig:RPHeadwind} that the LMC is infalling primarily face-on (in the $\hat{z}$-direction) for most of the past Gyr, until it turns towards edge-on in the last few hundred Myrs. The inclination and orbit we use again follow from \cite{Salem2015RAMMEDIUM}, who backward-orbit integrate the LMC - Milky Way system (neglecting the SMC) using the methods described in \cite{Besla2007}. As we will see, gas expulsion effected by ram pressure is quite sensitive to the LMC inclination during infall.

\subsection{The Star Formation History of the LMC}
\label{HZ2014}

Crucial for our study of outflow generation from the Clouds, we also have a sense of the star formation history of the Clouds, resolved fairly well in both space and time, by comparing observed and synthetic color-magnitude diagrams \cite{Harris2004SMC, Harris2009THECLOUD}. Within the framework of our isolated LMC simulations, for which we would like to focus on the interplay between outflow launching and ram pressure stripping, we study only the last Gyr of the LMC when ram pressure is non-negligible. During this recent time period, the star formation history of the LMC is also temporally well-resolved; the Harris and Zaritsky reconstruction is log-binned in time, so fluctuations in star formation are captured best within the past Gyr. As in B2018, then, we focus on the last Gyr of the LMC's orbit, but \emph{we now directly utilize aspects of the star formation history derived by \citet{Harris2009THECLOUD} to seed our outflow launching.}



\begin{figure}
\centering
\includegraphics[width = 0.49\textwidth]{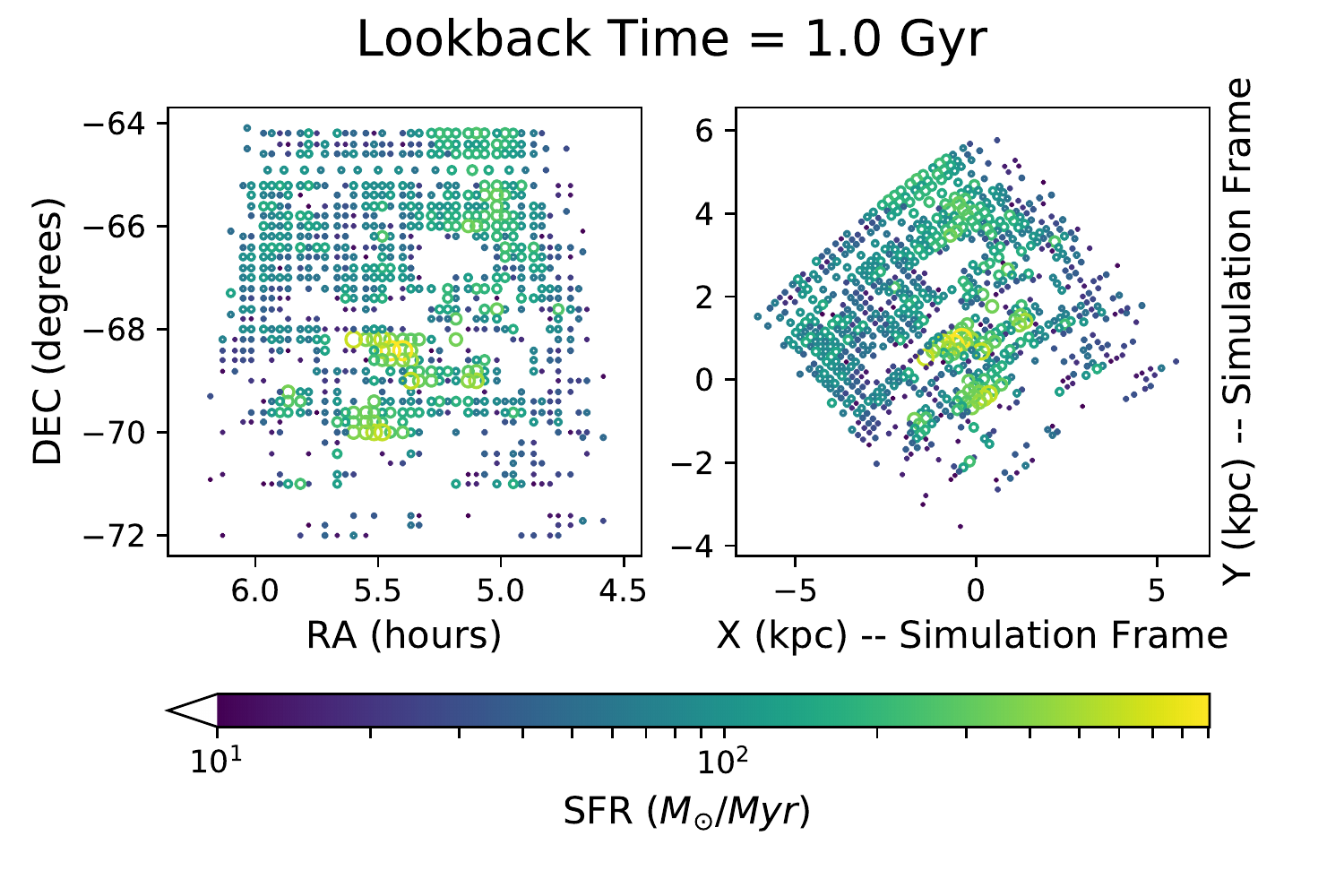}
\includegraphics[width = 0.49\textwidth]{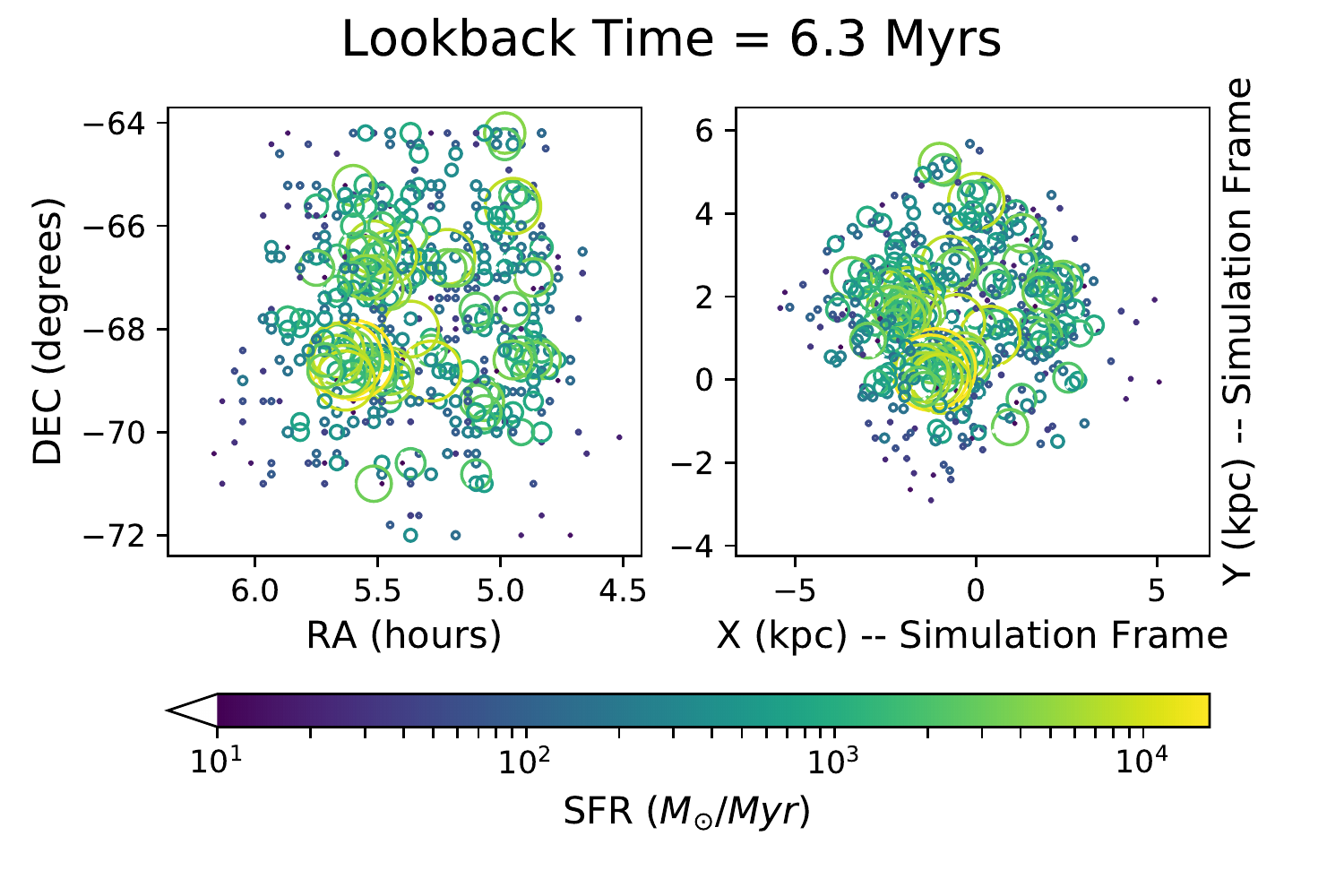}
\caption{Figure showing the SFRs from \cite{Harris2009THECLOUD} in RA-DEC and in simulation coordinates at two different time snapshots. Note the change in colorbar scale and change in areas of the circles, which represents the SFR of each region. Present-day (bottom) is dominated more by concentrated, highly-star forming regions than 1 Gyr ago (top), when star formation is more spread out. These circles are then spatially scrambled (not shown here) according to the algorithm described in Figure \ref{StarFormationGraphic} and represented as one or more active particles in FLASH.}
\label{RADECtoSimFrame}
\end{figure}

\section{Using the Derived Star Formation History of the LMC}
\label{sec:SFRHistory}
\begin{figure*}[t]
\label{StarFormationGraphic}
\centering
\includegraphics[width = 1.0\textwidth]{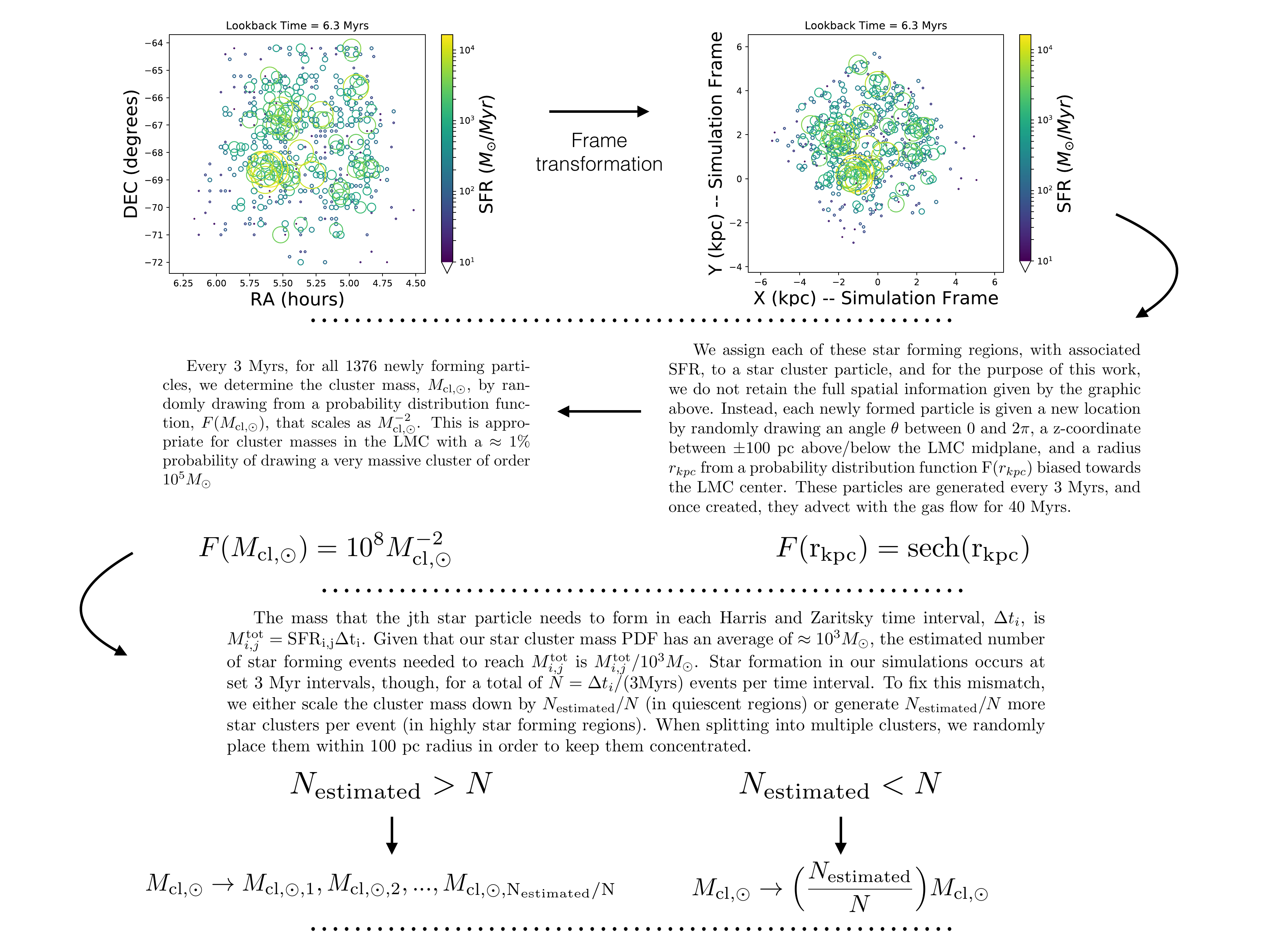}
\caption{Description of how we use the derived star formation history of the LMC given by \cite{Harris2009THECLOUD} to seed star cluster formation. Most importantly, we transform the star mass formed within each time bin (or the average star formation rate during that time bin) to a cluster mass. Every 3 Myrs, newly formed clusters, represented by active particles, are randomly placed within our simulation box according to our chosen radial distribution function and given a mass drawn from a mass distribution function motivated by observationally derived cluster masses in the LMC \citep{Glatt2010}. To ensure that the correct amount of star mass is formed (subject to a density threshold of $10^{-25} \rm g cm^{-3}$) at each step, the cluster mass is then either scaled down or broken into multiple (but still spatially clustered) star cluster particles. This method results in a bit of stochasticity while, on average, retaining the cumulative SFR given by \cite{Harris2009THECLOUD}. The decision to scale down the cluster mass or split into multiple clusters depends on the ratio $N_{\rm estimated}/N$, which reduces to $3 \times$ SFR/($10^{3} M_{\odot}/Myr$). For the snapshot shown above, the large SFRs of $\approx 10^{4} M_{\odot}/Myr$ are each split into $\approx 30$ cluster particles.}
\end{figure*}

\citet{Harris2009THECLOUD} give the star formation rate at 1376 positions in RA-DEC coordinates spread about the optical center of the LMC at $(\rm RA, \rm DEC) = (82.24 \degree ,-69.5 \degree)$. As described in the Appendix, we transform these positions to our simulation frame by first converting to the LMC frame and rotating by 100 degrees about the angular momentum axis. One can immediately see from Figure \ref{RADECtoSimFrame} that the grid defined by \citet{Harris2009THECLOUD} is biased towards the North (towards positive y in our simulation frame) relative to both the optical and kinematic centers of the disk. This skew is physical, however, as the LMC disk does extend to the North in a line pointing towards the Milky Way \citep{2011A&A...535A.115I}. This points to a gravitational interaction between the Milky Way and LMC as the cause of this skew, which presents some challenges in interpreting our simulations that start with an axisymmetric disk. 
While using the full amount of spatial and temporal star formation information is enticing, we acknowledge that star formation is complex and environment-dependent. Without tidal effects, in particular, we cannot re-create the observed morphology of the LMC beyond our axisymmetric initial condition; therefore, we choose to scramble the positions of the star-forming regions given in \citet{Harris2009THECLOUD}, while retaining more spatial information is left to future work. A more detailed graphic of how star cluster particles are generated in our simulations is given in Figure \ref{StarFormationGraphic}.

It's important here to talk about clustering of supernovae, its effects on outflow generation, and whether we properly capture clustering given the stochasticity of our star formation algorithm. This is prudent to consider given that idealized simulations suggest that clustered supernovae are far more efficient at driving observed outflows than random, homogeneous feedback (e.g. \cite{Fielding2017HowWinds, Fielding2018}). Highly clustered feedback was successful in \cite{Bustard2018}, where we injected purely thermal energy from a 30 Doradus-like cluster every 60 Myrs. Such a massive cluster was able to drive a localized fountain flow even though only thermal energy, subject to efficient radiative cooling, was deposited. While the actual distribution of clusters in the LMC is uniquely concentrated at present-day, we know this was not the case for the entirety of the LMC's infall. A large improvement can be made by utilizing the temporally and spatially reconstructed star formation history from \citet{Harris2009THECLOUD}.

First, note that we \emph{do} primarily retain the distribution of star cluster masses given in each time bin of the \citet{Harris2009THECLOUD} star formation reconstruction. That is, each circle in Figure \ref{RADECtoSimFrame} represents a single star cluster in our model. At every star formation time, separated by 3 Myrs by assumption, each cluster's position is randomly drawn according to the distribution functions given in Figure \ref{StarFormationGraphic}, but the amount of star mass formed at each step is motivated by the mass that the cluster needs to generate over that time bin. 

For example, for the time snapshot at a lookback time of 6.3 Myrs, one large star forming region (the lower left region including 30 Doradus) accounts for much of the star formation rate. The \citet{Harris2009THECLOUD} data is at such a resolution that this region is represented by multiple enhancements in SFR, including one very large SFR represented as a large circle in our graphic. When our simulation reads in this SFR snapshot, the clusters within this region are separated by our algorithm, but the largest circle $\emph{is}$ still present in our simulation. This circle falls within the parameter space where $N_{\rm estimated} > N$, so it is not one single, massive cluster, but instead a set of many particles, roughly conforming to the cluster mass distribution appropriate for the LMC. These particles are \emph{not} given random locations throughout the whole disk; instead, they are tied together within 100 pc of each other. This example shows that while our algorithm does not fully retain the level of clustering apparent from the \citet{Harris2009THECLOUD} data, it is partially retained. At earlier times (further in the past), the star formation rate over all 1376 positions is much more uniform than present-day, and this regularity is retained in our model despite the 1376 positions being randomized. Retaining various levels of spatial clustering, either in radial information only or in exact 2D position, is an intriguing exercise in outflows driven by supernovae in varying environments, which we leave for a future study.

For now, we note that these simulations, despite employing a density threshold for star formation, effectively inject energy at random locations, which is known to produce more energetic outflows than if supernovae are tied to the densest gas regions \citep{Walch2015, Simpson2016}, which is more realistic. One possible way to address this in the future is to form star particles through a more traditional pathway based on the ratio of cell mass to free-fall time or the frequently used prescription of \cite{Cen1992}. By then changing the star formation efficiency, one can try to match the cumulative star formation history of the LMC. This would more readily tie the star particles to the dense gas in which they should be forming and evolving, but it is not clear how to recreate the LMC's unique distribution of cluster masses that is important for outflow driving. As this star formation prescription is also more computationally expensive, given that self-gravity would need to be included, we leave this possibility to future work. 

Lastly, the smaller-scale effects of clustering are difficult to capture at our necessarily modest resolution, fiducially $\approx 78$ pc. While our algorithm decides to either scale down the mass or split up large SFR enhancements into a group of concentrated particles depending on the ratio $N_{\rm estimated}/N$, this should only matter at high resolution when overlapping shocks can be resolved. When $N_{\rm estimated} > N$, we split the cluster into multiple particles, but those particles are concentrated within a 100 pc radius that is comparable to our fiducial resolution. At higher resolution, energy injection from these split particles could result in a boost in asymptotic momentum scaling superlinearly with the number of explosions \citep{Gentry2017}. However, at our resolution, energy deposition from one massive particle and from many smaller but concentrated particles is effectively the same.

On smaller scales, feedback from each individual cluster particle also represents multiple overlapping supernova explosions. To reflect the momentum boost seen by \cite{Gentry2017}, \cite{Semenov2017}, for example, amplifies the momentum output per particle by a factor of 5 in their simulations. In our fiducial simulations, we do not include such a boost factor, but we have run one simulation without cosmic ray feedback with a factor of 5 boost in momentum deposition. Both simulations (with and without a boost) resulted in similar fountain flows far weaker than the outflows energized by cosmic rays.

\section{Computational Platform and Input Physics}
\label{methods}
Our computational tool of choice is the FLASH v4.2 magnetohydrodynamics (MHD) code \citep{FLASHRef}. Within this framework, we use the directionally unsplit staggered mesh solver \citep{2009JCoPh.228..952L, 2013JCoPh.243..269L}, which is based on a finite-volume, high-order Godunov scheme. This solver employs a constrained transport (CT) method to enforce the divergence free magnetic field condition. 

We also use an additional cosmic ray module that evolves cosmic rays as a second, relativistic fluid in addition to the usual thermal gas \citep{Yang2012,Ruszkowski2017}. Crucially, it includes a fluid approximation to a kinetic-scale instability, referred to as the cosmic ray streaming instability, that dominates the motion of the bulk cosmic ray population for $\approx$ GeV energy cosmic rays, which carry most of the momentum \citep{1969ApJ...156..445K, 1974ARA&A..12...71W, 2017PhPl...24e5402Z}. The resulting streaming transport is well-described by the ``self-confinement" model, in which the bulk cosmic ray population excites magnetic fluctuations if the cosmic ray drift velocity exceeds the local Alfv\'{e}n speed, $v_{A} = B/\sqrt{4\pi \rho}$. The cosmic rays then pitch-angle scatter off of these waves, confining the bulk cosmic ray population to flow down their pressure gradient, along the magnetic field direction, at the local Alfv\'{e}n speed. 

This transport differs from a standard diffusion process \citep{2017MNRAS.467..906W} more commonly implemented in hydrodynamic solvers; importantly, there is a transfer of energy (in the form of gas heating due to damping of the hydromagnetic waves excited by cosmic rays) between the cosmic ray and thermal gas populations. We refer to this cosmic ray energy loss as ``collisionless" because the energy transfer is mediated by magnetic waves instead of direct interactions between cosmic rays and ambient gas. 

If the cosmic rays do not self-excite their confining magnetic fluctuations and instead scatter off a turbulent cascade, which we refer to as the ``extrinsic turbulence" model \citep{2017PhPl...24e5402Z}, there is no transfer of energy from cosmic rays to the gas. Whether the cosmic rays are tightly locked to the gas (effectively the advection case) or diffuse through it depends on the amplitude of the turbulence. In our following simulations, we will explore cosmic ray driven outflows with both advection and streaming transport, assuming that a canonical 10\% of each supernova's energy is converted to cosmic ray energy through diffusive shock acceleration (see the Appendix for more details on our feedback implementation). 

For our streaming simulations, we additionally assume that all gas is fully ionized and streaming occurs at the Alfv\'{e}n speed, though cosmic ray propagation through partially neutral media may be highly super-Alfv\'{e}nic \citep{Farber2018}. This is less likely to be important in a dwarf galaxy than a Milky Way-mass or high-density starburst galaxy, where cosmic rays could free-stream through copious areas of partially neutral gas, but it is unclear how the combination of super-Alfv\'{e}nic streaming and cosmic ray collisional losses, both important in dense environments, affect the cosmic ray flux through the disk-halo interface. Future work comparing simulated vs observed gamma ray luminosities may inform our knowledge of cosmic ray transport (as in e.g. \cite{2017ApJ...847L..13P, Chan2019CRs}), and we plan to do this in the future.

Putting together the usual ideal MHD equations with the additional influence of cosmic rays, our simulations solve the following equations:

\begin{equation}
    \frac{\partial \rho }{\partial t} + \nabla \cdot (\rho \mathbf{u_{g}}) = 0
\end{equation}
\begin{equation}
    \frac{\partial \rho \mathbf{u_{g}} }{\partial t} + \nabla \cdot (\rho \mathbf{u_{g}} \mathbf{u_{g}} - \frac{\mathbf{B} \mathbf{B}}{4 \pi}) = \rho \mathbf{g} + \dot{p}_{SN}
\end{equation}
\begin{equation}
\frac{\partial B }{\partial t} - \nabla \times (\mathbf{u_{g}} \times \mathbf{B}) = 0
\end{equation}
\begin{equation}
\begin{split}
\frac{\partial e }{\partial t}  & + \nabla \cdot \left[(e+p_{tot}) \mathbf{u_{g}} - \frac{\mathbf{B}(\mathbf{B} \cdot \mathbf{u_{g}})}{4 \pi} \right] = \rho \mathbf{u_{g}} \cdot \mathbf{g} \\
& - \nabla \cdot \mathbf{F_{c}} - C + H_{c} + H_{SN} + \Gamma_{th}
\end{split}
\end{equation}
\begin{equation}
    \frac{\partial e_{c} }{\partial t} + \nabla \cdot (e_{c} \mathbf{u_{g}}) = -p_{c} \nabla \cdot \mathbf{u_{g}} - H_{c} + H_{SN} - \nabla \cdot \mathbf{F_{c}} - \Lambda_{c}
\end{equation}
where $\rho$ is the gas density, $\mathbf{u_{g}}$ is the gas velocity, $\mathbf{B}$ is the magnetic field, $p_{tot} = (\gamma_{g} - 1) e_{g} + (\gamma_{c} - 1) e_{c} + \mathbf{B}^{2} / 8 \pi$ is the total pressure, and $e = 0.5 \rho \mathbf{u_{g}}^{2} + e_{g} + e_{c} + \mathbf{B}^{2}/8 \pi$ is the total energy density: the sum of kinetic energy density, gas energy density ($e_{g}$), cosmic ray energy density ($e_{c}$), and magnetic energy density. Note that the cosmic ray adiabatic index is $\gamma_{c} = 4/3$, while the gas adiabatic index is $\gamma_{g} = 5/3$. The following terms are due to cosmic ray streaming: $\mathbf{F_{c}}$ is the cosmic ray flux due to streaming, where we assume in this work that the streaming speed is $v_{A}$ but must be approximated following the regularization method \citep{2009arXiv0909.5426S} implemented by \cite{Ruszkowski2017} in FLASH (see Appendix for more details); $H_{c}$ is the heating of the gas due to damping of waves generated by the streaming instability, which one can show goes as $v_{A} \cdot \nabla P_{c}$ (e.g. \cite{2017PhPl...24e5402Z}). $\dot{p}_{SN}$ and $H_{SN}$ encode the momentum and heating from supernovae, which is described in detail in the Appendix. C and H are radiative cooling and heating terms for the thermal gas. $\Lambda_{c} = \Lambda_{\rm hadr} + \Lambda_{\rm coul}$ represents the collisional energy loss due to cosmic ray hadronic and Coulomb interactions. $\Gamma_{th} = \Lambda_{\rm hadr}/6 + \Lambda_{\rm coul}$ is the associated energy gain for the thermal gas. All energy from cosmic ray Coulomb interactions is thermalized, heating the background gas, while only 1/6 of the energy from hadronic interactions is thermalized. The rest of the hadronic energy loss escapes as gamma-rays. We use the equations of \cite{Enblin2007, Pfrommer2017} for the Coulomb and hadronic loss terms:

\begin{equation}
    \Lambda_{\rm coul} = -2.78 \times 10^{-16} \left(\frac{n_{e}}{\rm cm^{-3}}\right) \left(\frac{e_{c}}{\rm erg cm^{-3}}\right) \rm erg s^{-1} cm^{-3}
\end{equation}
\begin{equation}
    \Lambda_{\rm hadr} = -7.44 \times 10^{-16} \left(\frac{n_{e}}{\rm cm^{-3}}\right) \left(\frac{e_{c}}{\rm erg cm^{-3}}\right) \rm erg s^{-1} cm^{-3}
\end{equation}
where $n_{e}$ is the electron number density tabulated in \cite{Wiersma2009ThePlasmas} as a function of density and temperature assuming photoionization equilibrium with the metagalactic UV background \citep{2012ApJ...746..125H}.


\begin{table*}[]
    \centering
    \begin{tabular}{c c c c c c c }
        LMC Gas Mass & B Field Strength & Cosmic Rays & Ejected Mass (from disk) & Ejected Mass (from sphere)  \\
        \hline
        \hline
        \emph{Ram Pressure Only} & & & & & & \\
        \hline
        {\bf low} & weak & no & $5.39 \times 10^{7} M_{\odot}$ & $2.65 \times 10^{7} M_{\odot}$ &&\\
        {\bf high} & strong & no &  $6.05 \times 10^{7} M_{\odot}$ & $2.32 \times 10^{7} M_{\odot}$ &\\
        \hline
        \emph{Feedback Only} & & & & &&\\
        \hline
        {\bf low} & weak & no  & $4.78 \times 10^{6} M_{\odot}$  &  $2.07 \times 10^{6} M_{\odot}$&\\
        {\bf low} & weak & yes & $2.56 \times 10^{8} M_{\odot}$  & $6.04 \times 10^{7} M_{\odot}$&\\
        high & strong & no & 0 & 0 &\\
        {\bf high} & strong & yes  & $5.10 \times 10^{8} M_{\odot}$ & $1.63 \times 10^{7} M_{\odot}$&\\
        {\bf high} & strong & yes (w/streaming)  & $7.80 \times 10^{7} M_{\odot}$   & $3.62 \times 10^{6} M_{\odot}$&\\
        \hline
        \emph{Feedback + Ram Pressure} & & & & &&\\
        \hline
        high & strong & yes & $4.50 \times 10^{8} M_{\odot}$ & $2.63 \times 10^{7} M_{\odot}$& \\
        high & strong & yes (w/streaming) & $1.06 \times 10^{8} M_{\odot}$ & $1.69 \times 10^{7} M_{\odot}$ & \\
        \hline

    \end{tabular}
    \caption{Table of simulations run, broken down by LMC gas mass of $M_{\rm gas} = 5 \times 10^{8} M_{\odot}$ (low) and $M_{\rm gas} = 10^{9} M_{\odot}$ (high), magnetic field strengths peaking at $1 \mu G$ (weak) and $4 \mu G$ (strong), and whether cosmic ray feedback is included. The final two columns show the amount of gas expelled from the \emph{disk}, defined as the gas with ISM tracer fraction $> 0.01$ outside a disk of radius 13 kpc and height 1.7 kpc above and below the midplane, and the amount of ISM expelled outside a sphere of radius 13 kpc, which is the cutoff radius of the initial disk.}
    \label{tab:table}
\end{table*}

\subsection{Radiative Cooling}
We also include radiative cooling in our simulations assuming the gas is in photoionization equilibrium with the metagalactic UV background \citep{2012ApJ...746..125H}. For comparison to observations, including a photoionizing background is crucial as low-temperature gas, especially at low densities, is significantly affected and may be quite far from collisional equilibrium. The LMC and Magellanic Stream are highly ionized \citep{Barger2017}. New data from the Wisconsin H-Alpha Mapper (WHAM) suggests that the ionized mass fraction of the LMC and its extended halo is between $\approx 50$ and 75$\%$ \citep{Brianna2019AAS}. In addition to the $5 \times 10^{8} M_{\odot}$ of neutral gas within the central 4 kpc radius of the LMC, this ionized component pushes the total gas mass towards $10^{9} M_{\odot}$ or greater. Comparing our simulations to observations of neutral and ionized hydrogen, then, is much more accurate if we include photoionization.

The cooling function we utilize is a tabulated function of density and temperature from \citet{Wiersma2009ThePlasmas}, and the equation of state of the gas is updated accordingly at each timestep, as well, based on the gas ionization state tabulated as a function of density and internal energy. Subcycling is utilized to resolve the cooling time. A temperature floor of 300 K is included. 

The cooling rate or heating rate (depending on density - temperature regime) is calculated by adding the contribution from hydrogen and helium to the contribution from metals scaled by the metallicity. We neglect for now any ionizing photons from the Milky Way or the LMC itself, as well as non-equilibrium effects; however, we note that photoionization, similar to non-equilibrium cooling, extends high ionization states down to lower temperatures, which lessens the difference between non-equilibrium and equilibrium cooling \citep{2013MNRAS.434.1043O}. Additionally, most stripped gas is already fairly low temperature (hence, it is not actively radiating very much), and it is likely cooling mostly due to adiabatic expansion. Non-equilibrium effects may be important, though, in the supernova-heated outflows, where gas cools down from temperatures greater than $10^{6}$ K or so. 

The metallicity is tracked by a tracer fluid that is initially set to $0.3 Z_{\odot}$ for gas of densities greater than the initial background halo density, and set to $0.01 Z_{\odot}$ for lower density gas (the Milky Way halo). Mass ejected from supernovae is enriched to $2.0 Z_{\odot}$ in our simulations, but this gas quickly mixes with the ISM, keeping the ISM metallicity close to $0.3 Z_{\odot}$. While the surrounding halo is initially set to be pristine in order to suppress cooling-driven condensation of the Milky Way halo onto the LMC disk, outflows that break out of the disk after $\approx 400$ Myrs do pollute the surrounding medium with a higher metallicity halo. This halo gas is then more subject to radiative cooling, especially as subsequent outflows compress the existing halo gas, and can precipitate onto the LMC disk or the Trailing Stream; a higher metallicity for the ambient Milky Way halo may amplify this condensation and will be included in future work. Indeed, O VI is detected out to the virial radius \citep{Tumlinson2011,Johnson2015}, and a number of Milky Way analogs seem to have elevated metallicities $0.3 Z_{\odot}$ or greater \citep{Stern2016,Prochaska2017,Bregman2018}. The interaction of this ambient gas with satellite galaxies may trigger precipitation (as seen in e.g. \cite{Wright2019, 2019ApJ...882..156H}), and the Magellanic System may be a local example of this.




\section{Results}
\label{sec:results}

Table \ref{tab:table} shows the set of simulations that we will analyze in this section. Each simulation captured the most recent Gyr of LMC infall. Note that the simulation snapshots are time-stamped with \emph{simulation time}, not lookback time. Present day is represented by a simulation time of 1.0 Gyr and a lookback time of 0.0. Each of the simulations with feedback included had a base resolution of 312 pc and a maximum resolution of 78 pc, except for the low gas mass LMC outflow simulations, for which the base resolution was 1250 pc to save computational expense. Refinement was done based on density, with a threshold for refinement of $10^{-26} \rm g cm^{-3}$. The ram pressure only simulations, since they do not have such high temperature, timestep-limiting gas, are run with a maximum resolution of 39 pc. Each simulation without ram pressure was run on a (40 kpc)$^{3}$ box, with the LMC placed at the grid center, while each simulation with ram pressure was run on a (60 kpc)$^{3}$ box with the LMC centered at (-10 kpc, -10 kpc, -10 kpc). Resolution studies varying maximum and base resolutions were carried out, as well, and the results are presented in the Appendix.

\subsection{Results with Only Feedback}

Outflows with and without CRs from our low and high gas mass LMC disks are launched using our star formation prescription and feedback implementation outlined in the Appendix. We track the mass expelled from the disk, defined as a cylinder of radius 13 kpc and height 1.7 kpc above and below the midplane, and from a sphere of radius 13 kpc. The results are given in Table \ref{tab:table}. While in the purely thermal case (no cosmic rays) much of the deposited thermal energy is lost to radiative cooling, the cosmic ray population sustains a pressure gradient capable of blowing out a far more powerful outflow. In the high gas mass case, while neither the thermally driven nor cosmic ray driven outflows significantly expel gas beyond a sphere of radius 13 kpc, the cosmic ray driven outflow unbinds more than $5 \times 10^{8} M_{\odot}$ from the disk region. Gas is driven even further out of the gravitational potential well in the low gas mass LMC simulation, but less gas is expelled from the disk overall because there is half as much gas to begin with (see Table \ref{tab:table}). We list the mass expelled from the high gas mass LMC with thermal winds as 0 because the initial ISM mass outside of our defined disk region is actually higher than at present-day, signalling that thermal winds couldn't puff the disk back up to its pre-collapse height.

\begin{figure}
    \centering
    \includegraphics[width = 0.45\textwidth]{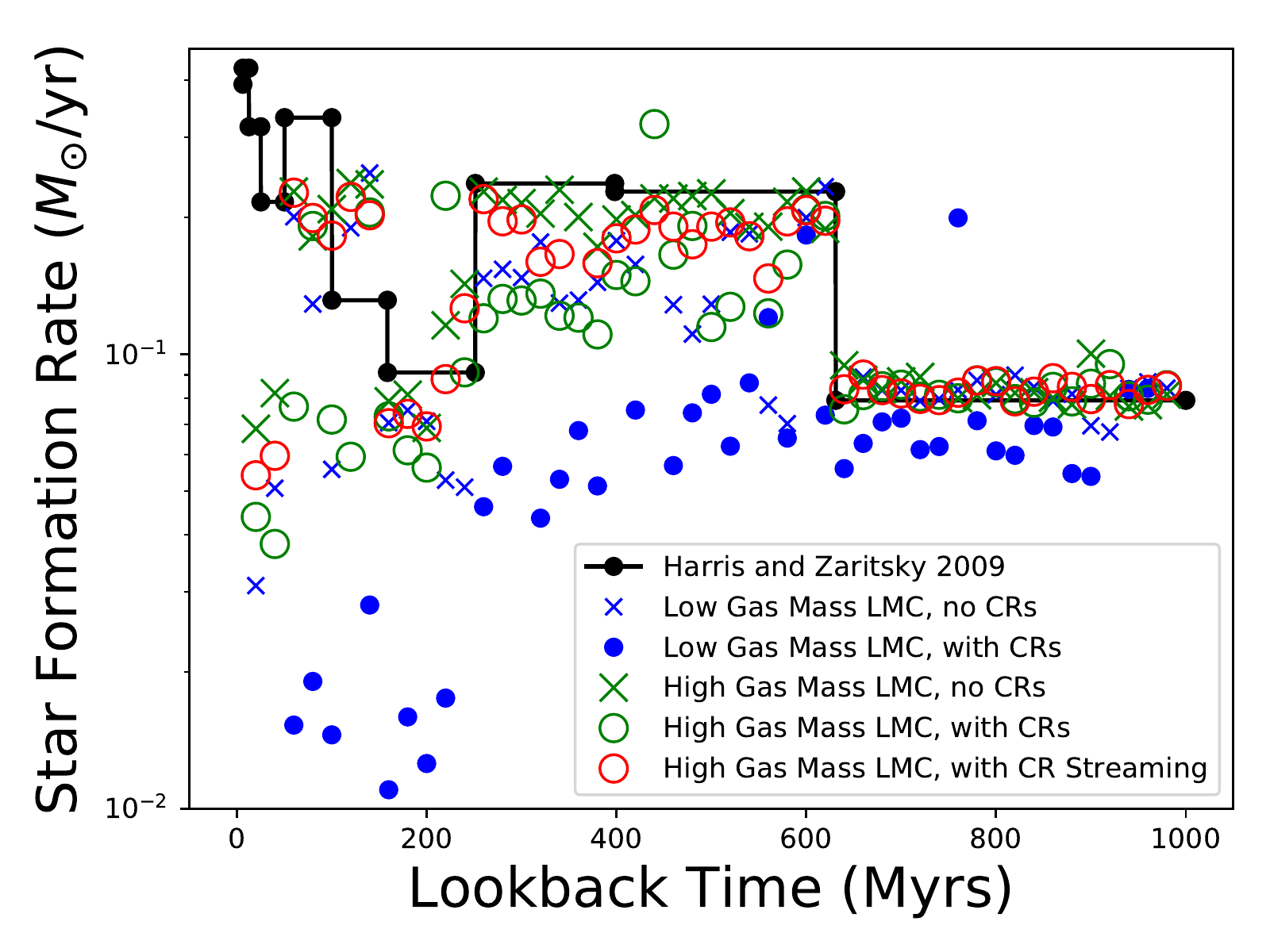}
    \caption{Star formation rate (SFR) as a function of time for simulations compared to intended SFR from \cite{Harris2009THECLOUD}. Low gas mass LMC simulations, especially with cosmic ray driven outflows, severely underestimate the intended SFR because they expel much of the dense gas from the disk. A higher gas mass LMC provides more dense gas for star formation and weighs down the disk, making it harder for outflows to break out. The simulation with cosmic ray streaming and collisional losses gives a decent match, at least following the trend as well as simulations without cosmic rays, which blow out almost no gas.}
    \label{fig:SFR}
\end{figure}

Figure \ref{fig:SFR} shows the actual SFR for each of our simulations compared to the intended SFR. While the low gas mass LMC with thermally driven outflows matches the intended SFR fairly well (until present-day, when it falls short), the cosmic ray driven outflow blows out a significant amount of dense gas, leaving only a few cells that satisfy the density threshold for star formation. This causes the actual SFR to fall short of the intended value even early on in our simulation. This is rectified somewhat by increasing the LMC gas mass, which leads to a much closer match between the actual and intended SFRs for our CR driven outflows. The thermally driven outflow, especially, matches the intended SFR very well because almost no dense gas is lofted above the disk. This gives us a first-order constraint on our simulations, suggesting that either 1) the high gas mass LMC is a more appropriate setup; 2) our star formation prescription is too simplistic, falling short because we throw out all star particles that cannot form in their pre-determined location, when, in fact, there may be dense star-forming gas elsewhere in the disk; or 3) cosmic ray driven outflows are too strong when we do not account for energy losses. This last point is of particular interest, as cosmic rays additionally diffuse or stream along magnetic field lines. In the advection picture, cosmic rays are well-trapped within the disk at early times, thereby generating a steep pressure gradient. Only after the wind is driven, cosmic rays can advect with the outflow and escape the disk. Indeed, we find in our advection-only simulations that the average cosmic ray pressure even within the disk at present-day is an order of magnitude higher than gas pressure, likely too high to be realistic. 

\begin{figure}
\centering
\includegraphics[width = 0.4\textwidth]{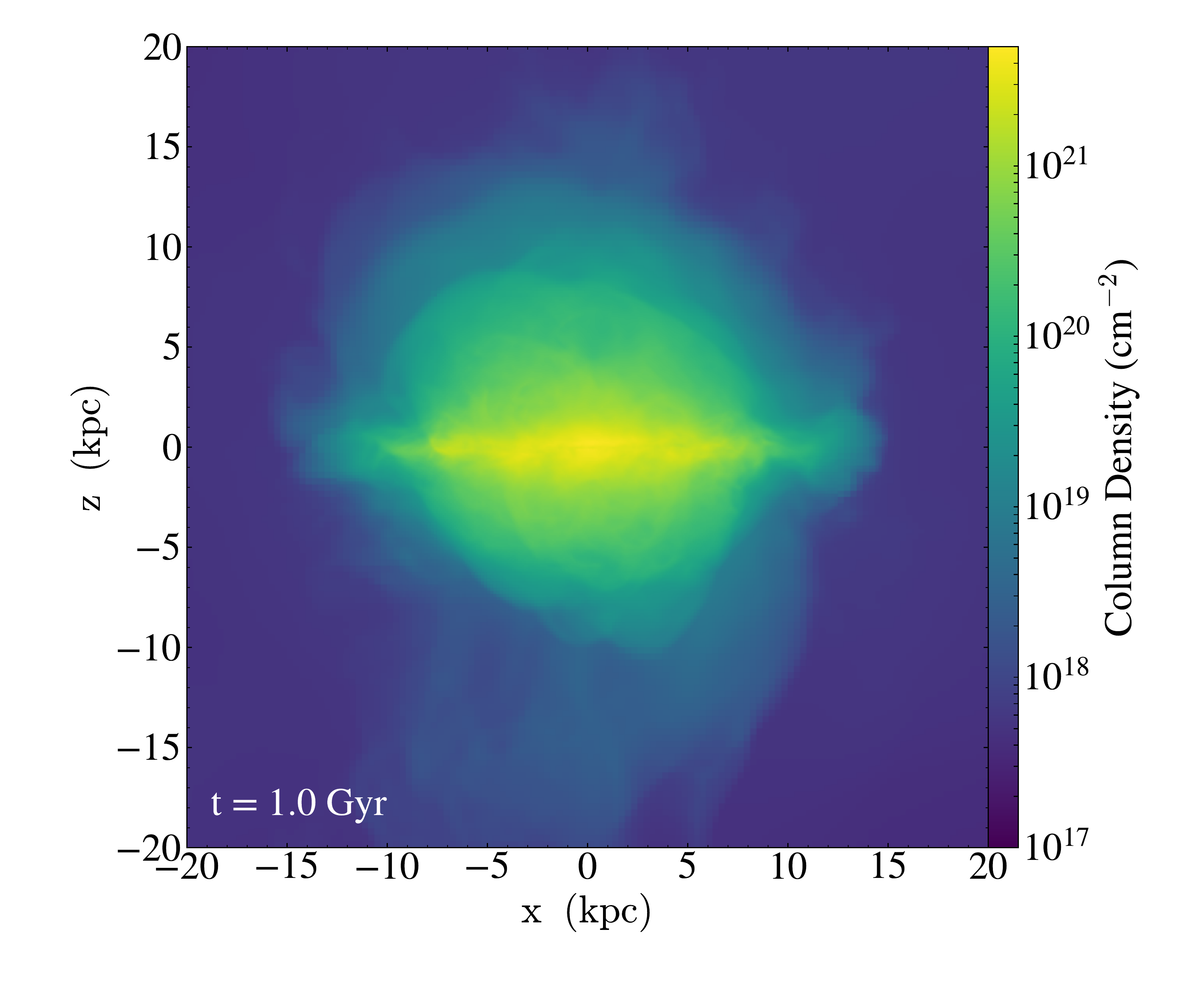}
\includegraphics[width = 0.4\textwidth]{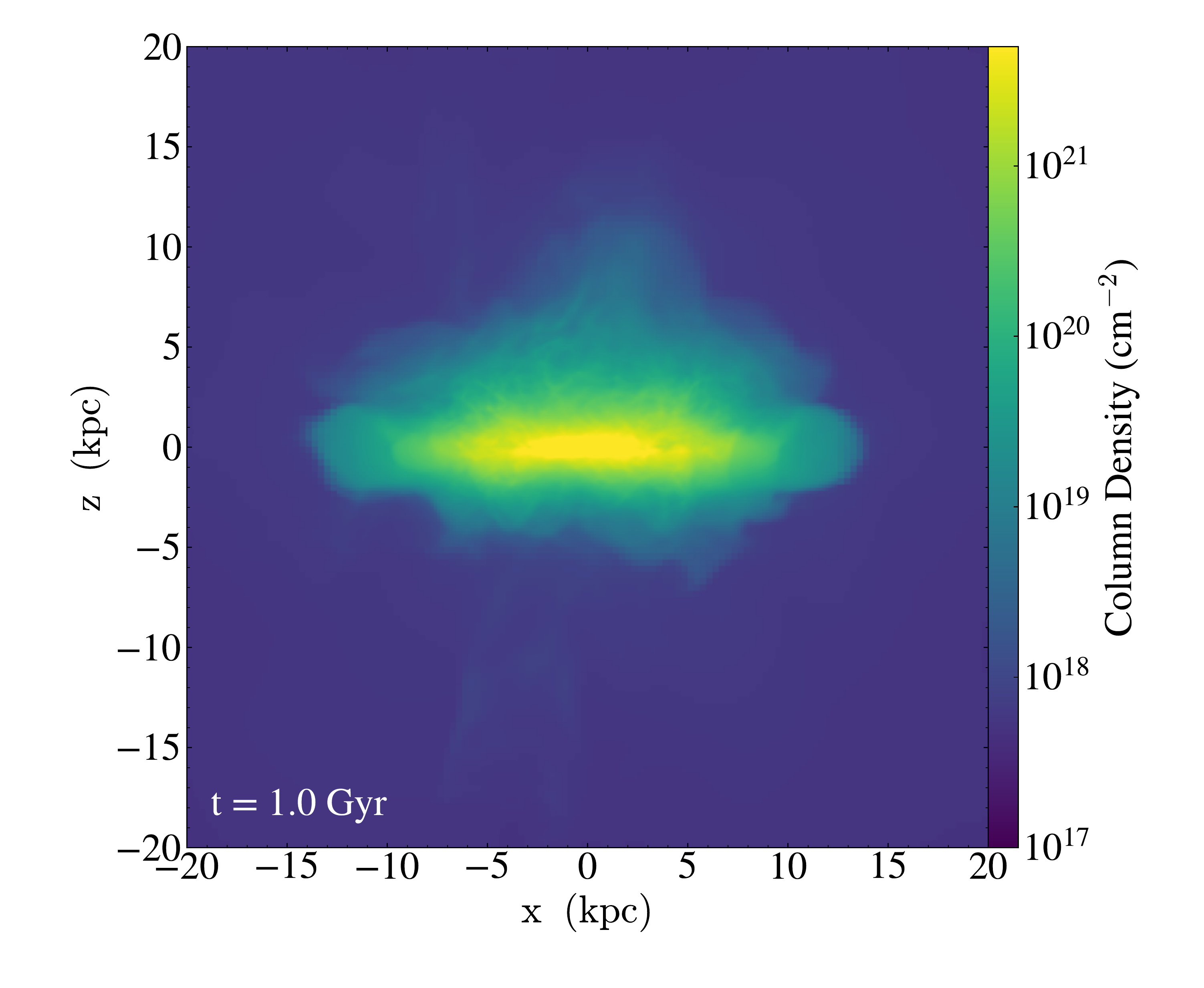}
\caption{Edge-on projection of total density for the high gas mass LMC with winds driven by cosmic rays with pure advection (top) and streaming plus collisional losses (bottom). Streaming and collisional energy transfer from the cosmic ray population to the thermal gas decreases the overall pressure gradient as thermal gas is susceptible to strong radiative cooling in the disk. This results in a weaker outflow than a pure advection case without any cosmic ray energy losses.}
\label{CRWind_lowmass_highmass}
\end{figure}

Analyzing the impact of cosmic ray streaming and collisional losses will be the subject of a forthcoming paper (Bustard et al. 2019, in prep), but we present one preliminary simulation with streaming and collisional losses. The resulting density projections edge-on are shown for our high gas mass LMC simulations with advecting and, additionally, streaming cosmic rays in Figure \ref{CRWind_lowmass_highmass}. The purely thermal feedback case is not shown, as no gas noticeably breaks out of the disk. With the more realistic cosmic ray treatment, some of the cosmic ray pressure is sapped by collisional and also collisionless losses due to streaming. A fraction of this energy heats the thermal gas, but this gained energy is now susceptible to radiative cooling. The net result is less energy available to drive the outflow and far less mass expelled from the disk ($7.80 \times 10^{7} M_{\odot}$ compared to $5.10 \times 10^{8} M_{\odot}$). This is reflected in Figure \ref{CRWind_lowmass_highmass}, which shows a much weaker outflow compared to the cosmic ray advection case without energy losses. Because less dense gas is expelled from the disk, the SFR also increases and actually becomes similar to the low gas mass LMC with no CRs (Figure \ref{fig:SFR}). 

That streaming results in a smaller outflow than advection seems to challenge other recent simulations, which tend to show that cosmic ray transport (either by diffusion or streaming) actually drives stronger outflows than advection, which simply puffs up the disk and suppresses star formation (e.g. \cite{2012MNRAS.423.2374U, Simpson2016, Ruszkowski2017}). However, we note that our simulations don't represent a like-to-like comparison, primarily because collisional losses are fairly significant in our streaming simulations, while they are not included in our advection simulations. This extra energy loss, a large portion of which goes into gamma-ray emission instead of thermal heating, decreases the outflow energy source and explains much of why these simulations give a weaker outflow. 

Beyond this, we note that most published simulations assume a Milky Way mass galaxy (though see e.g. \cite{Chan2019CRs, Hopkins2019CRs}), whereas the LMC is less massive and hence has a shallower gravitational potential well. The combination of thermal pressure and cosmic ray pressure locked to the gas may be sufficient to drive outflows, then, whereas in Milky Way mass galaxies, cosmic ray transport is necessary as it accelerates the Parker instability \citep{Heintz2018, Heintz2019} and redistributes cosmic ray pressure to greater heights, providing an additional driving mechanism \emph{outside} of the disk region \citep{2014MNRAS.437.3312S, Ruszkowski2017, Mao2018}. Further exploration of cosmic ray wind driving across a range of galaxy masses is needed to put our results in proper context. 
There are also differences in feedback implementation. For instance, we highlight the differences between our purposely imposed SFR and the star formation prescriptions used in other galaxy-scale cosmic ray driven outflow simulations, which tie star particles to only the dense gas regions through a prescription such as \cite{Cen1992}. This may be a source of the discrepancy as our advection and streaming simulations only vary by a small factor in SFR. In the simulations of e.g. \cite{Ruszkowski2017}, the cosmic ray advection case drives an initial mass flux but then feedback mostly shuts off, while the cosmic ray streaming case sustains a larger SFR, meaning more energy and momentum is available to drive an outflow. One may also attribute some of these discrepancies to varying topologies of the simulated ISM, resulting from differences in supernova placement, clustering, etc. The ISM structure then affects cosmic ray propagation and the effectiveness that advecting vs streaming cosmic rays have on driving outflows. 

\begin{figure*}
    \centering
    \includegraphics[width = 0.30\textwidth]{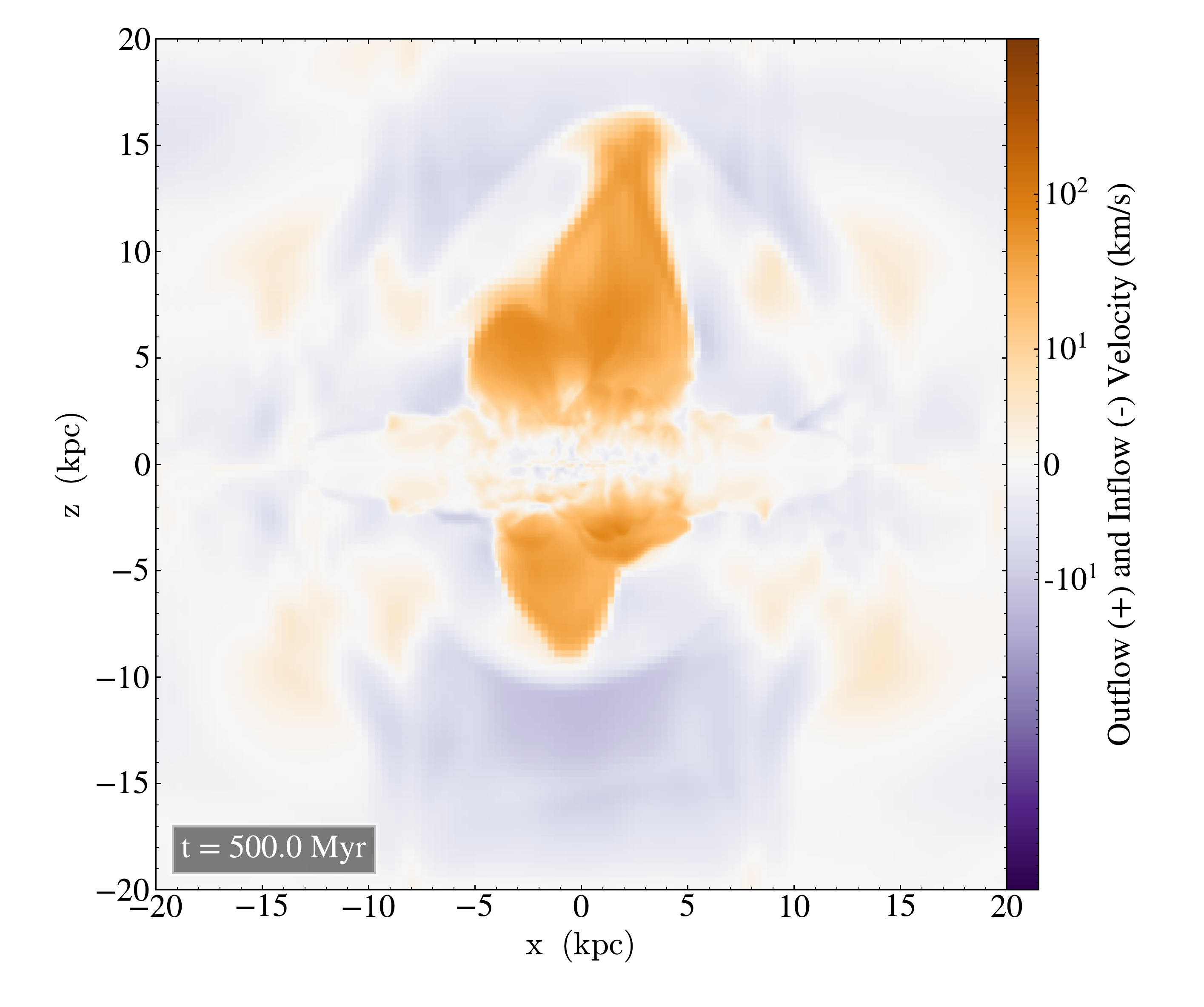}
    \includegraphics[width = 0.30\textwidth]{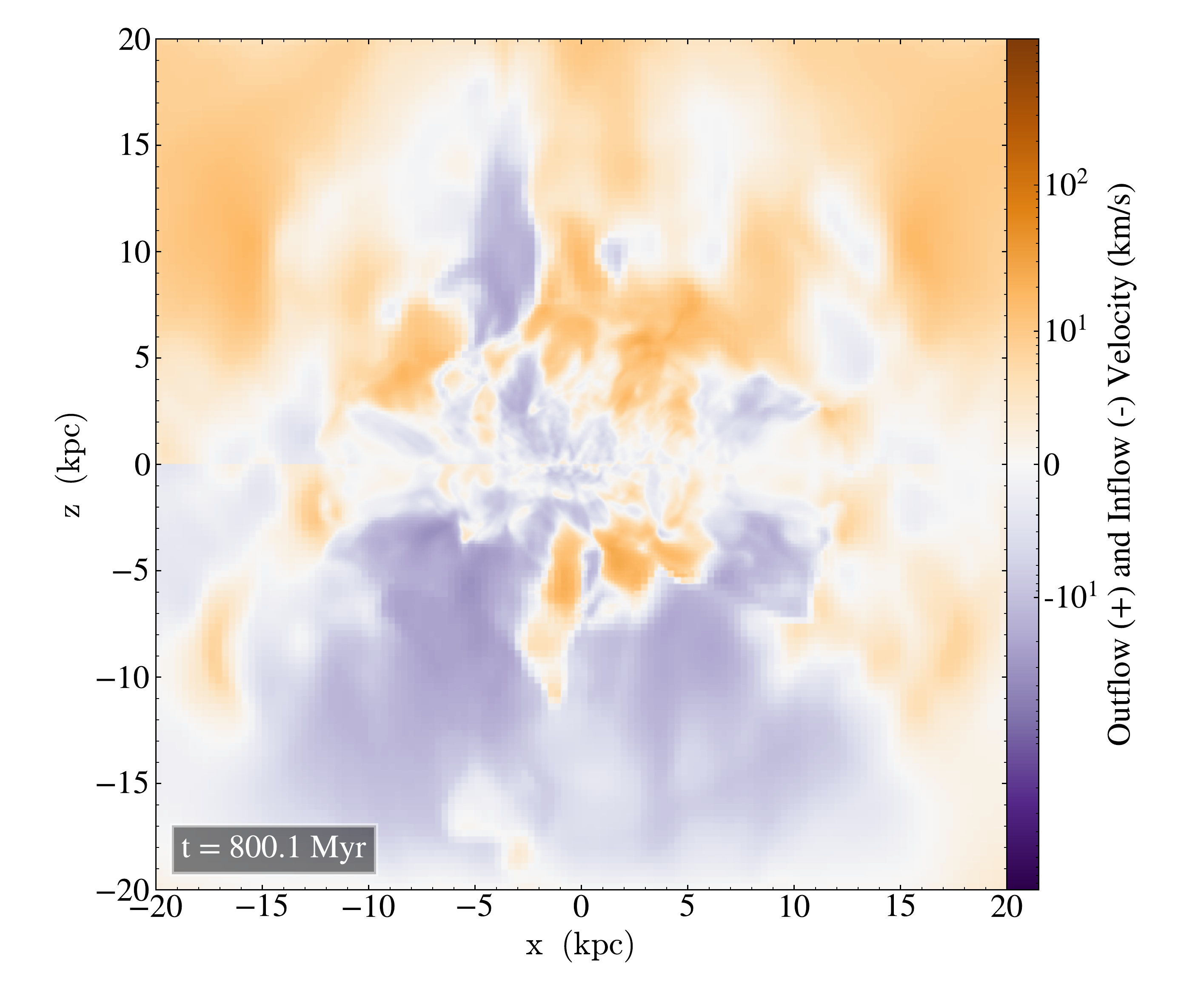}
    \includegraphics[width = 0.30\textwidth]{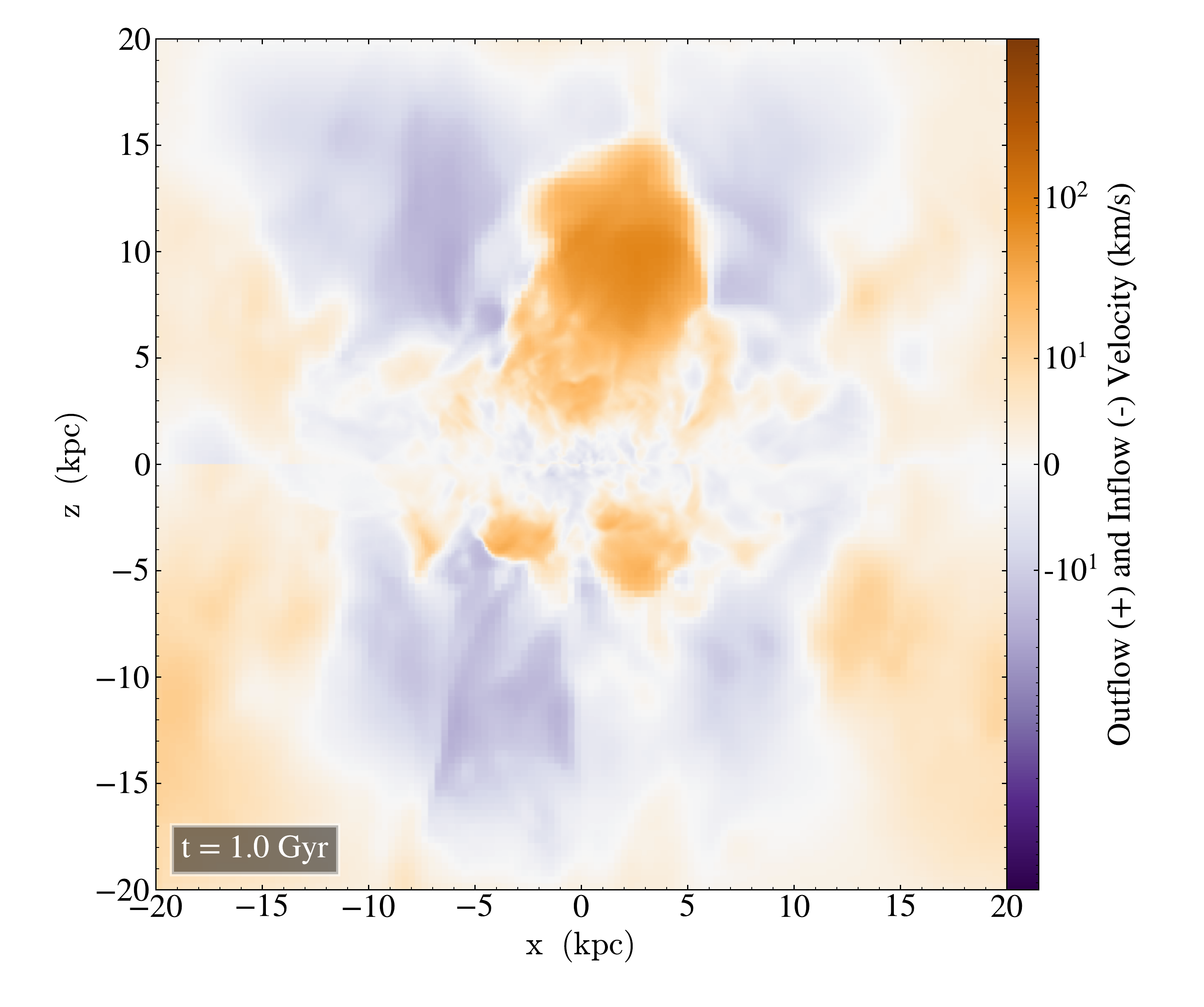}
    \caption{Edge-on, density-weighted projections of vertical velocity, both outflow and inflow, for the cosmic ray driven outflow (no ram pressure) from the high gas mass LMC with streaming and collisional losses. The timeseries of snapshots gives an indication of how bursty the outflows are, as large outflow events with velocities greater than a few hundred km/s occur when the star formation rate is high (the left and right panels), while infall is more predominant while the star formation is decreased (middle panel).}
    \label{fig:velocity_streaming}
\end{figure*}

\begin{figure}
    \centering
    \includegraphics[width = 0.45\textwidth]{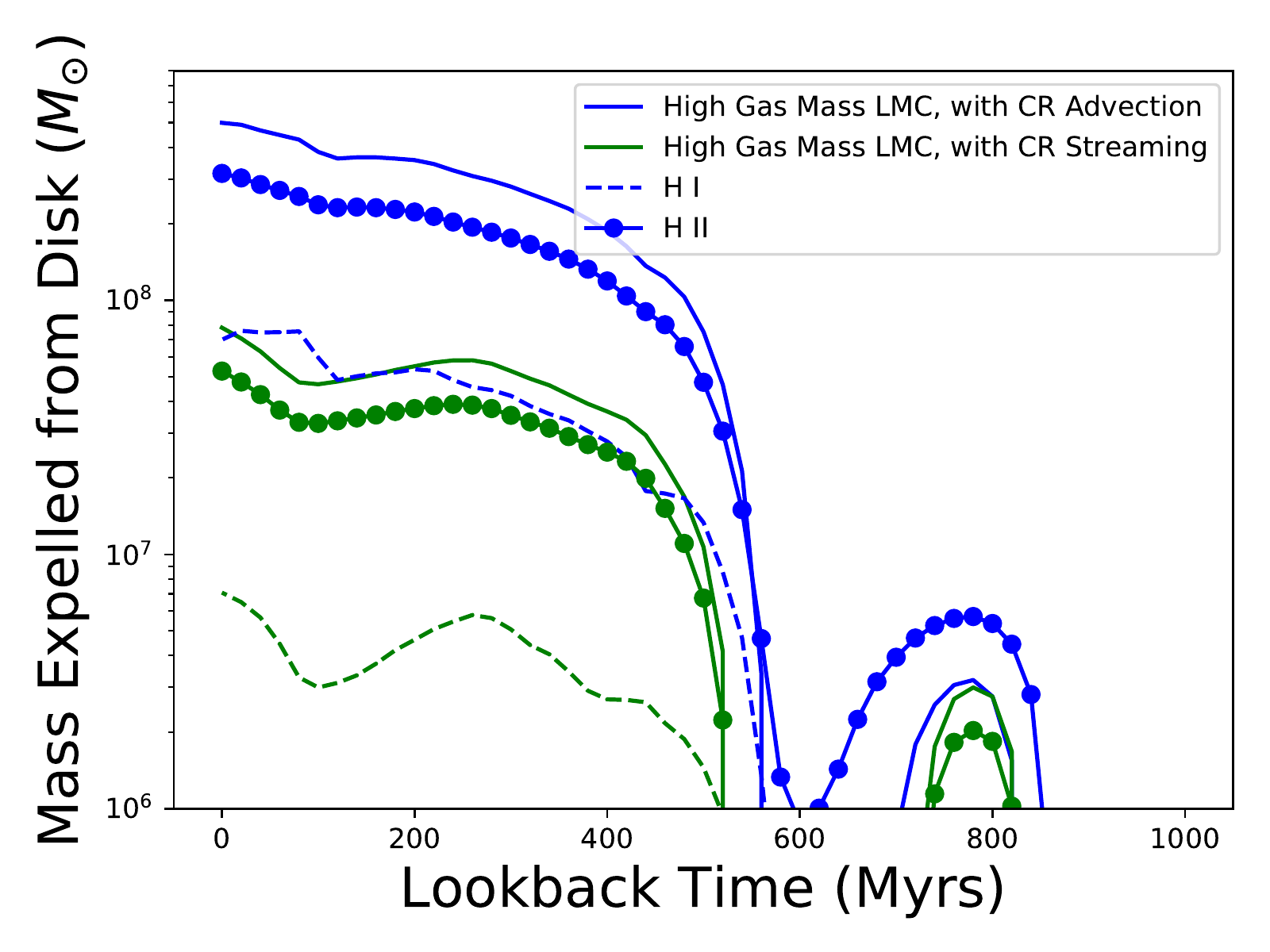}
    \caption{For simulations with feedback but no ram pressure, this shows the change in total mass and mass of neutral and ionized hydrogen measured outside a disk of radius 13 kpc and height 1.7 kpc (5 times the initial scale height) for two different cosmic ray treatments. At lookback times $\approx 800$ Myrs, note that the change in total mass is less than the change in H II mass. This is because the H I mass actually decreases when the initial puffy disk collapses, lowering both the total and neutral gas mass outside the disk, while the ionized gas mass sees a slight increase. The highest outflow mass flux occurs between 400 and 600 Myrs ago, when the SFR is highest. Another large mass expulsion episode is occurring within 100 Myrs of present-day in each simulation, too, as it follows another jump in the SFR. Almost all mass expelled is ionized hydrogen, with only a small amount of neutral hydrogen present. }
    \label{fig:massExpelled}
\end{figure}

For our simulations specific to the LMC, the best test of realism will come from detailed comparison to LMC observations, such as a comparison to gamma-ray observations, neutral and ionized hydrogen maps, and comparison to the best outflow estimates from \cite{Barger2016}. Figure \ref{fig:velocity_streaming} shows edge-on, density-weighted velocity projections of the high gas mass LMC with cosmic ray streaming. The three different time snapshots illuminate how bursty outflows are in these simulations, with large expulsion episodes followed by primarily quiescent periods where gas inflow rather than outflow dominates. These trends track the SFR very well, as seen if one compares the SFR of Figure \ref{fig:SFR} to the time-varying mass expelled from the disk shown in Figure \ref{fig:massExpelled}. A jump in the SFR 600 Myrs ago drives a large mass flux into the halo, which slows down for a brief period before another increase in SFR drives a large burst within the last 200 Myrs. As shown by Figure \ref{fig:massExpelled}, this trend is true for both the streaming and advection simulations, with the advection case expelling much more gas in the initial outburst. The ionization state of expelled gas is decomposed using the Trident package \citep{TridentRef}, which assumes, as we do, that the gas is in photoionization equilibrium with the metagalactic UV background. Expelled gas in both simulations is primarily ionized, as the diffuse gas is efficiently ionized by the background radiation field. This supports recent observations showing significant ionized hydrogen in the LMC halo and Trailing Stream, but the exact ionization state is sensitive to feedback implementation and resolution. 

Another useful comparison is to gamma ray emission, as that specifically gives us a handle on cosmic ray production and transport appropriate for the LMC. This will be the focus of future work, but we note that the preliminary simulated gamma-ray luminosity of our streaming simulation seems to overshoot the present-day estimated limit and is very close to calorimetric, meaning that a large fraction of cosmic ray energy is lost to hadronic and Coulomb collisions. This supports the conclusion that collisions can account for the decreased mass flux compared to the advection case. This also suggests that, even accounting for streaming at the Alfv\'{e}n velocity, which allows cosmic rays to somewhat escape collisions in the dense disk regions, the cosmic ray population is overproducing gamma-rays and, hence, losing more energy than we would expect. More efficient cosmic ray escape from e.g. super-Alfv\'{e}nic streaming is well-motivated \citep{Farber2018}, something that we will consider in future work and which has support from other recent simulations of cosmic ray driven winds from dwarf galaxies \citep{Chan2019CRs}.


\subsection{Including Ram Pressure}
\begin{figure*}
\centering
\includegraphics[width = 0.3\textwidth]{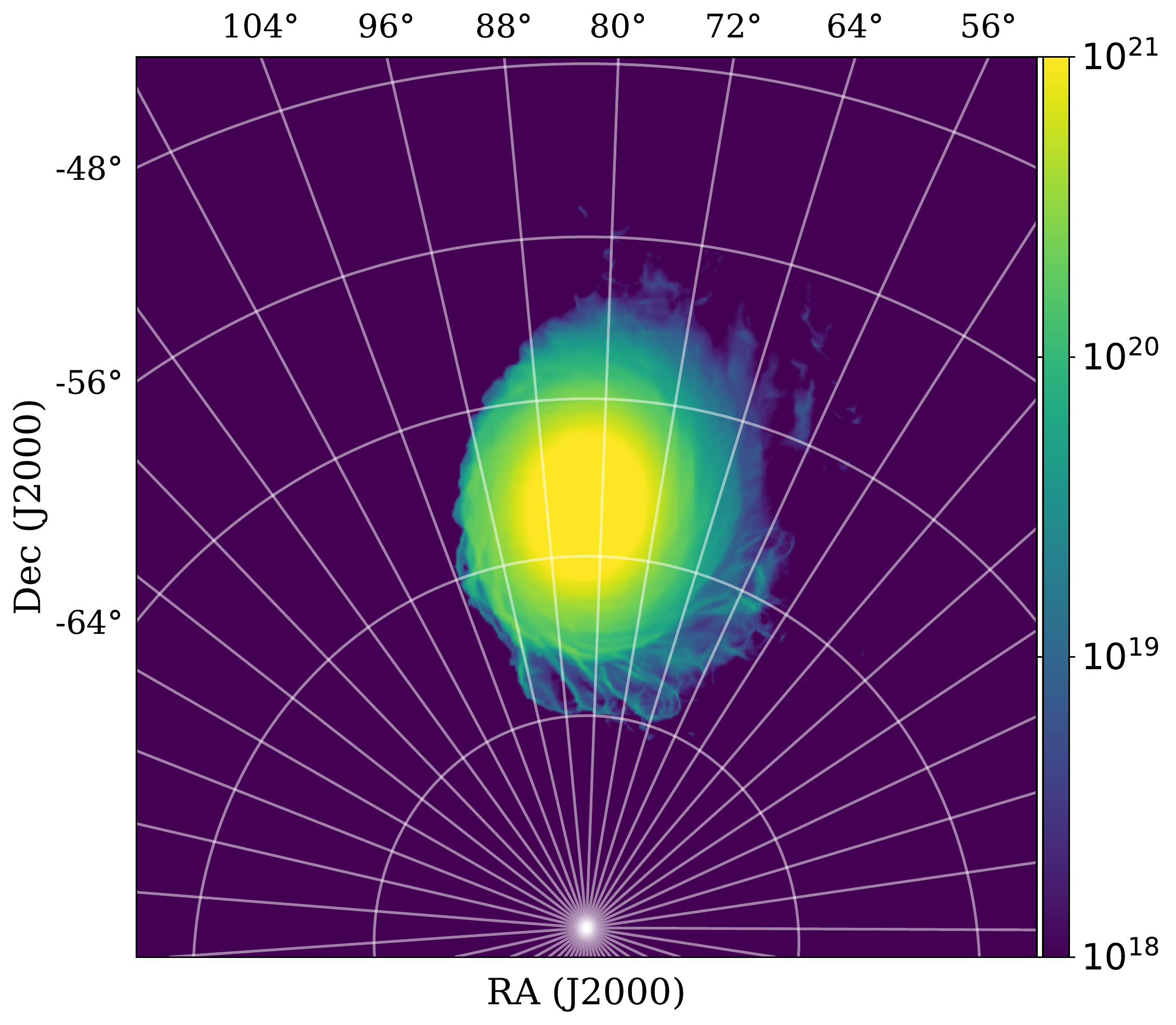}
\includegraphics[width = 0.30\textwidth]{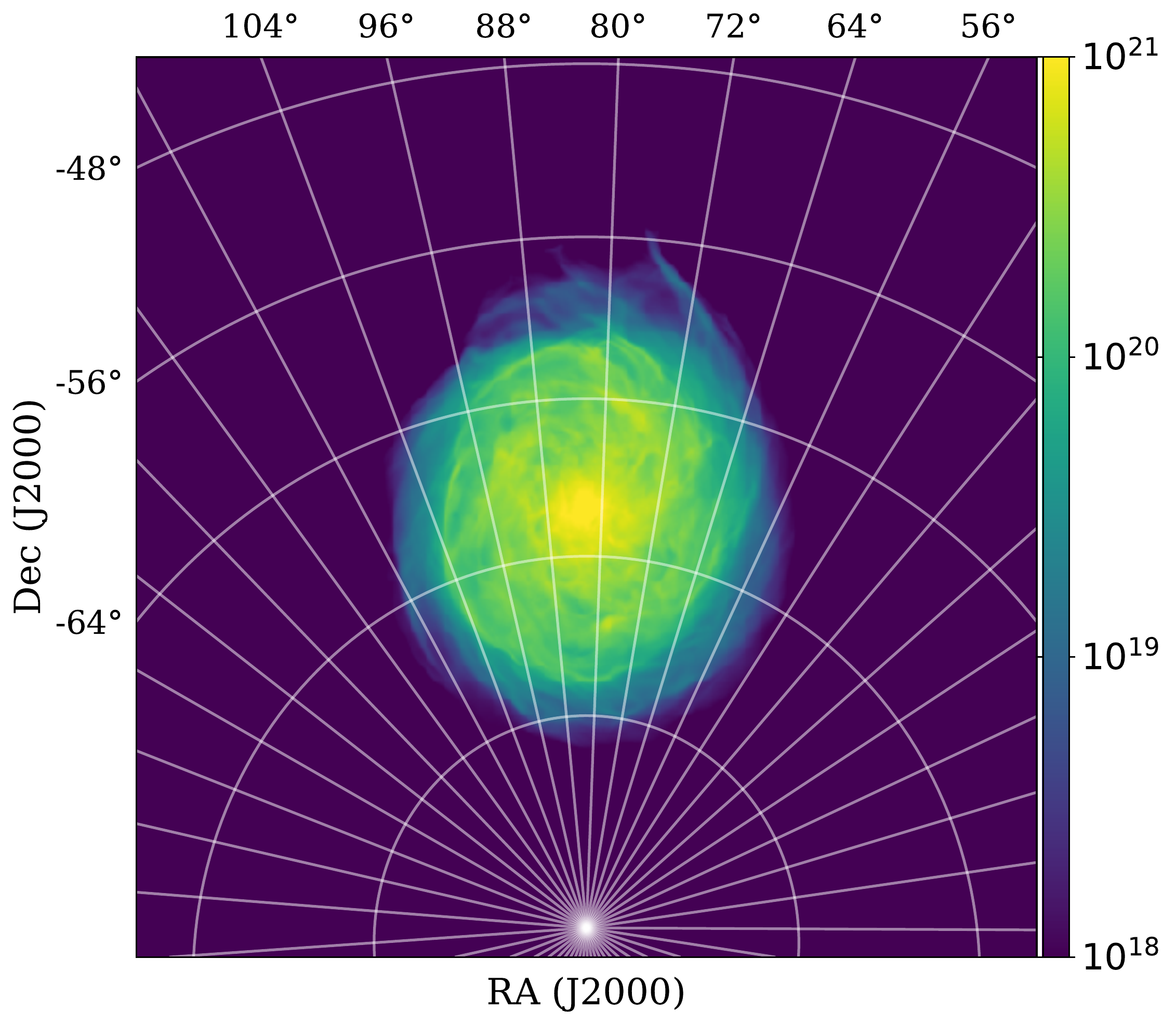}
\includegraphics[width = 0.3\textwidth]{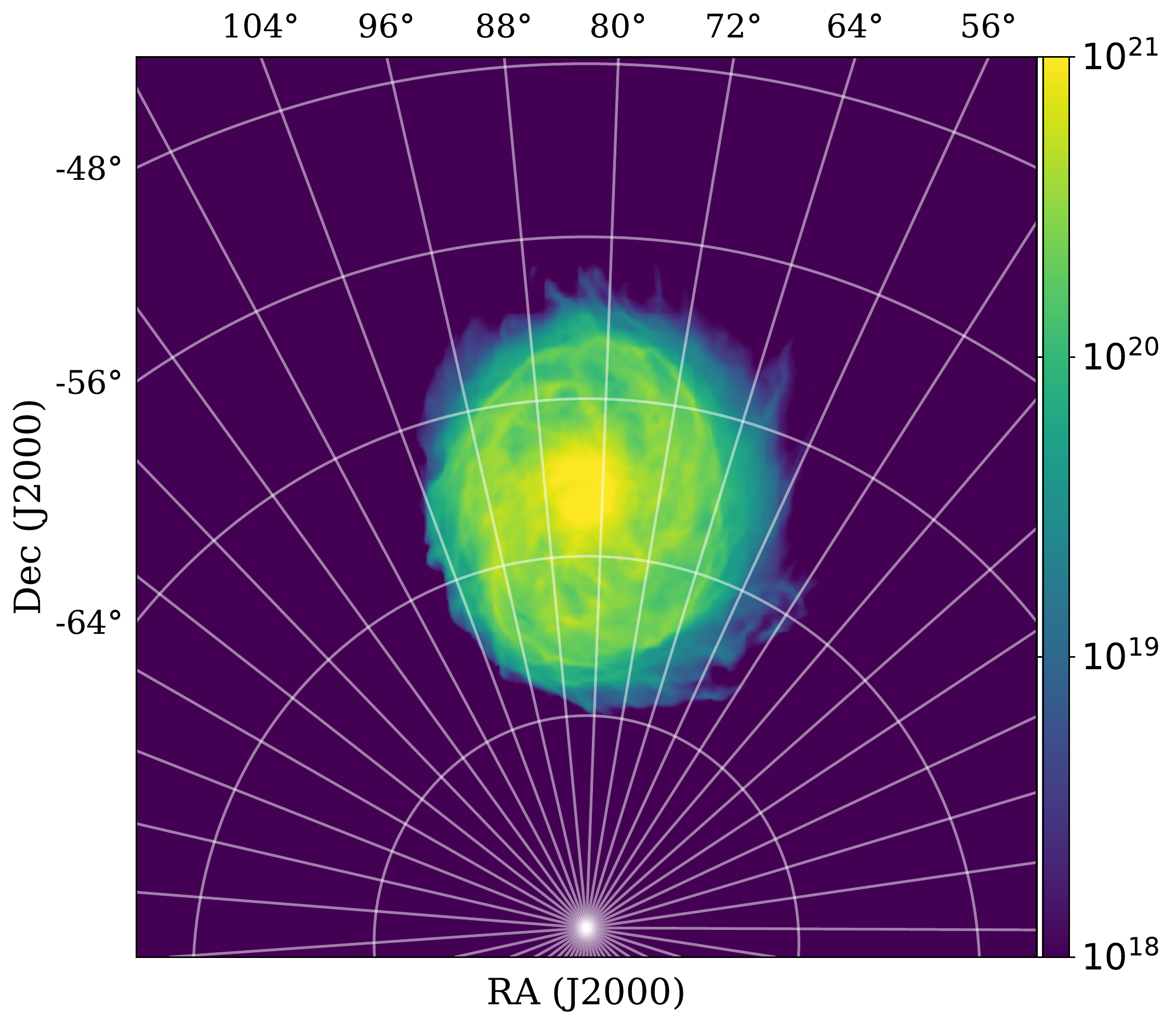}

\includegraphics[width = 0.30\textwidth]{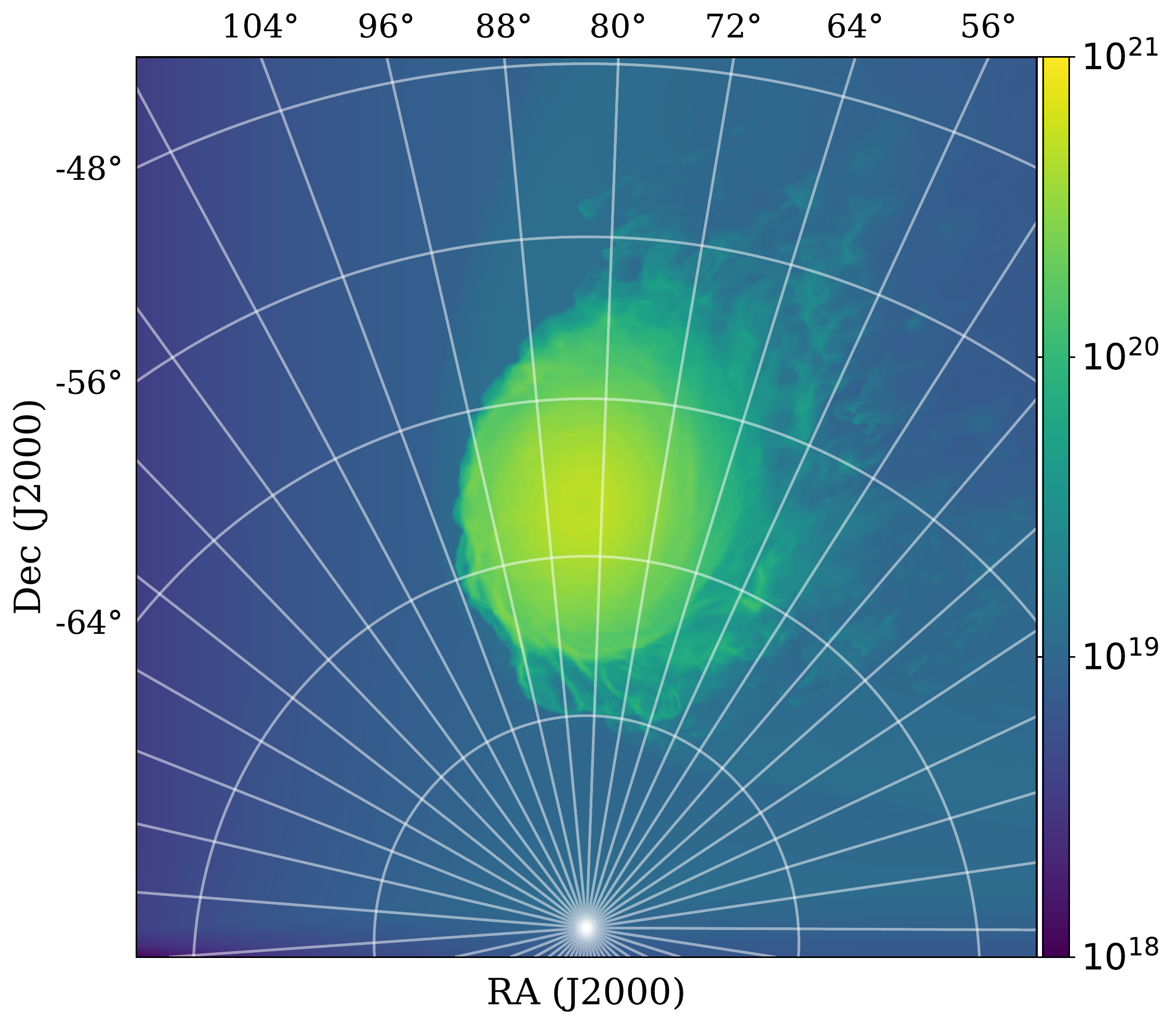}
\includegraphics[width = 0.30\textwidth]{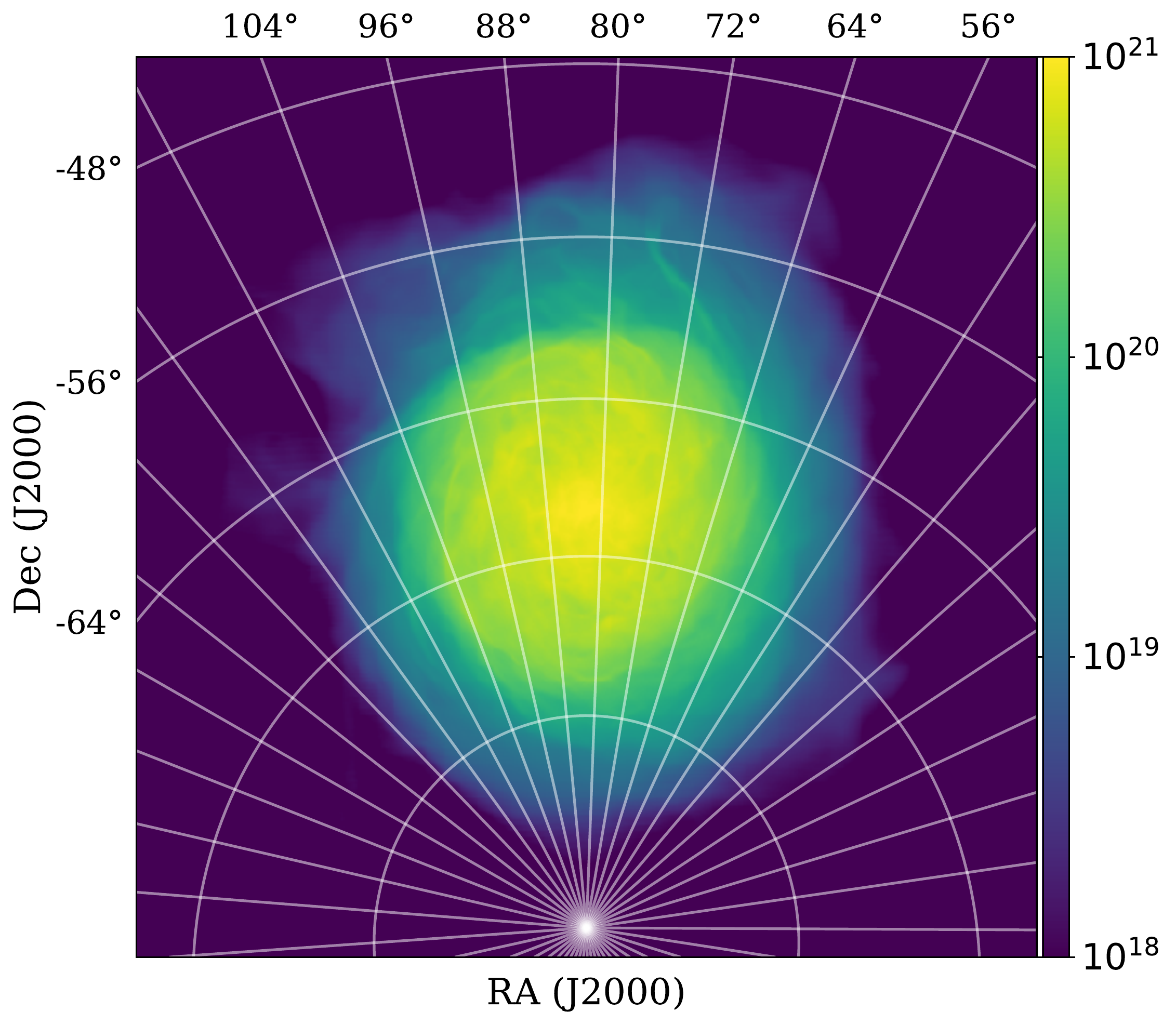}
\includegraphics[width = 0.30\textwidth]{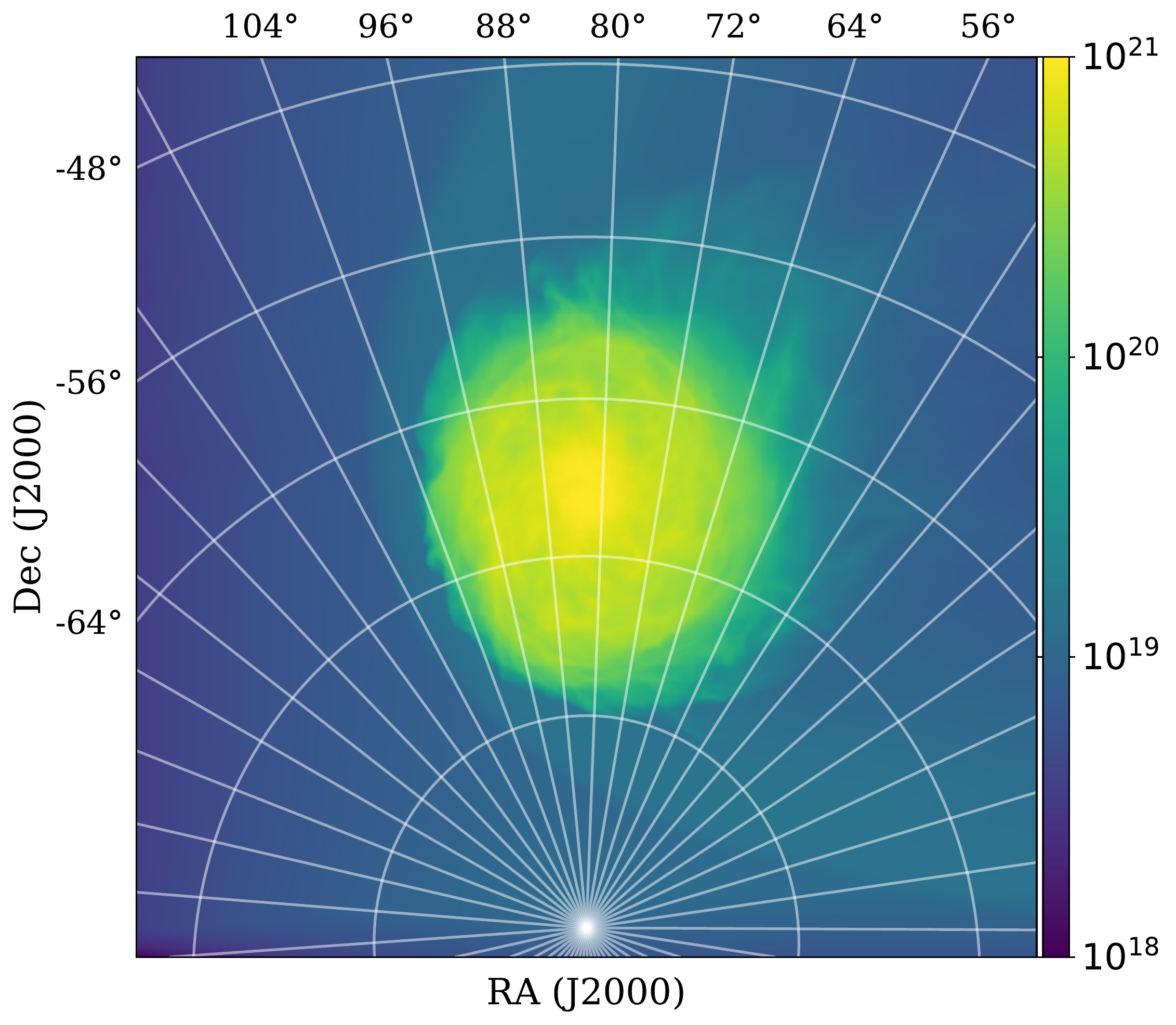}
\caption{Projected H I (top) and H II (bottom) column densities of the high gas mass LMC at present day. \emph{Left column:} Just ram pressure stripping; \emph{Middle column:} Just cosmic ray (advection) driven outflows; \emph{Right column:} Ram pressure plus cosmic ray (advection) driven outflows. Ram pressure stripped gas appears primarily as ionized hydrogen, with almost no visible tail present in neutral hydrogen. Overall, ram pressure is very ineffective alone. With feedback alone, strong cosmic ray driven outflows develop a hazy ionized hydrogen halo extending beyond the initial disk radius. With ram pressure included, this halo gets compressed along the leading edge (lower left of disk) but shows only a negligible tail structure (upper right of disk) as gas is primarily retained in the potential well on the far-side of the disk. }
\label{MockObservationFigure}
\end{figure*}

Beginning with ram pressure alone, we present synthetic H I and H II column densities in Figure \ref{MockObservationFigure} after 1 Gyr for our high gas mass LMC. The low gas mass LMC simulation (not shown) looks qualitatively very similar. Some stripped material, primarily ionized, protrudes from the upper right portion of the disk. Table \ref{tab:table} shows that the amount of gas expelled in both the low gas mass and high gas mass cases are quite similar and insignificant compared to the total mass of the Stream. This is consistent with the findings of \cite{Salem2015RAMMEDIUM}, as it should be since our galaxy setup and ram pressure inflow follows \cite{Salem2015RAMMEDIUM} very closely. 


We next model the combination of ram pressure and outflows from the LMC. While we have seen from our simulations with just outflows that only a negligible amount of gas is expelled beyond the LMC sphere of radius 13 kpc, we aim to see here whether the gas unbound from the LMC disk by outflows can be swept away by ram pressure, as was the case in B2018 for even small fountain flows and an edge-on ram pressure. We will focus on the cosmic ray (advection) driven outflow from the high gas mass LMC, which optimistically displaced greater than $5 \times 10^{8} M_{\odot}$ of ISM gas into the LMC halo, as well as the smaller, more realistic outflow driven by streaming cosmic rays with energy losses.

\begin{figure*}
    \centering
    \includegraphics[width = 0.4\textwidth]{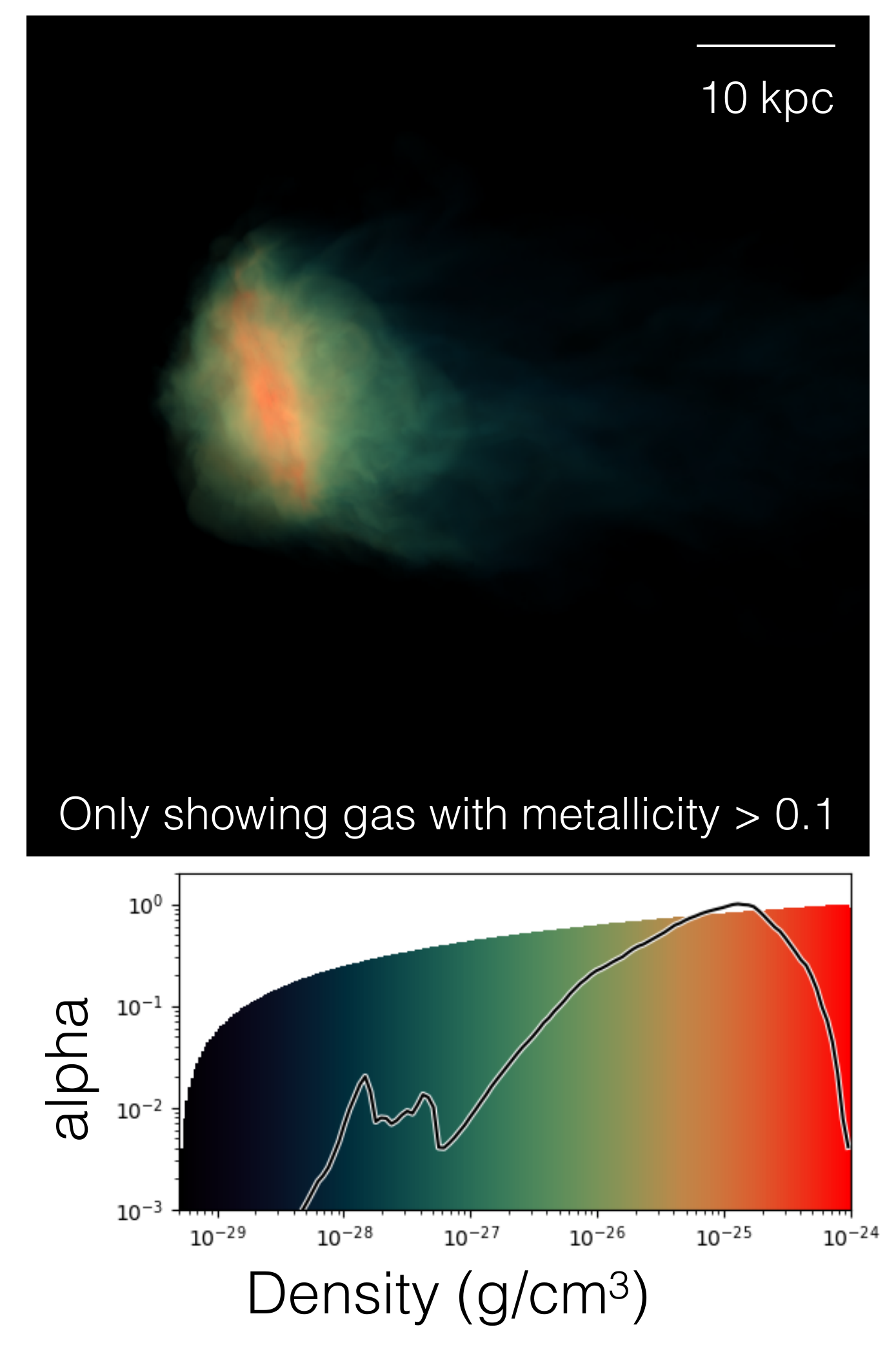}
     \includegraphics[width = 0.4\textwidth]{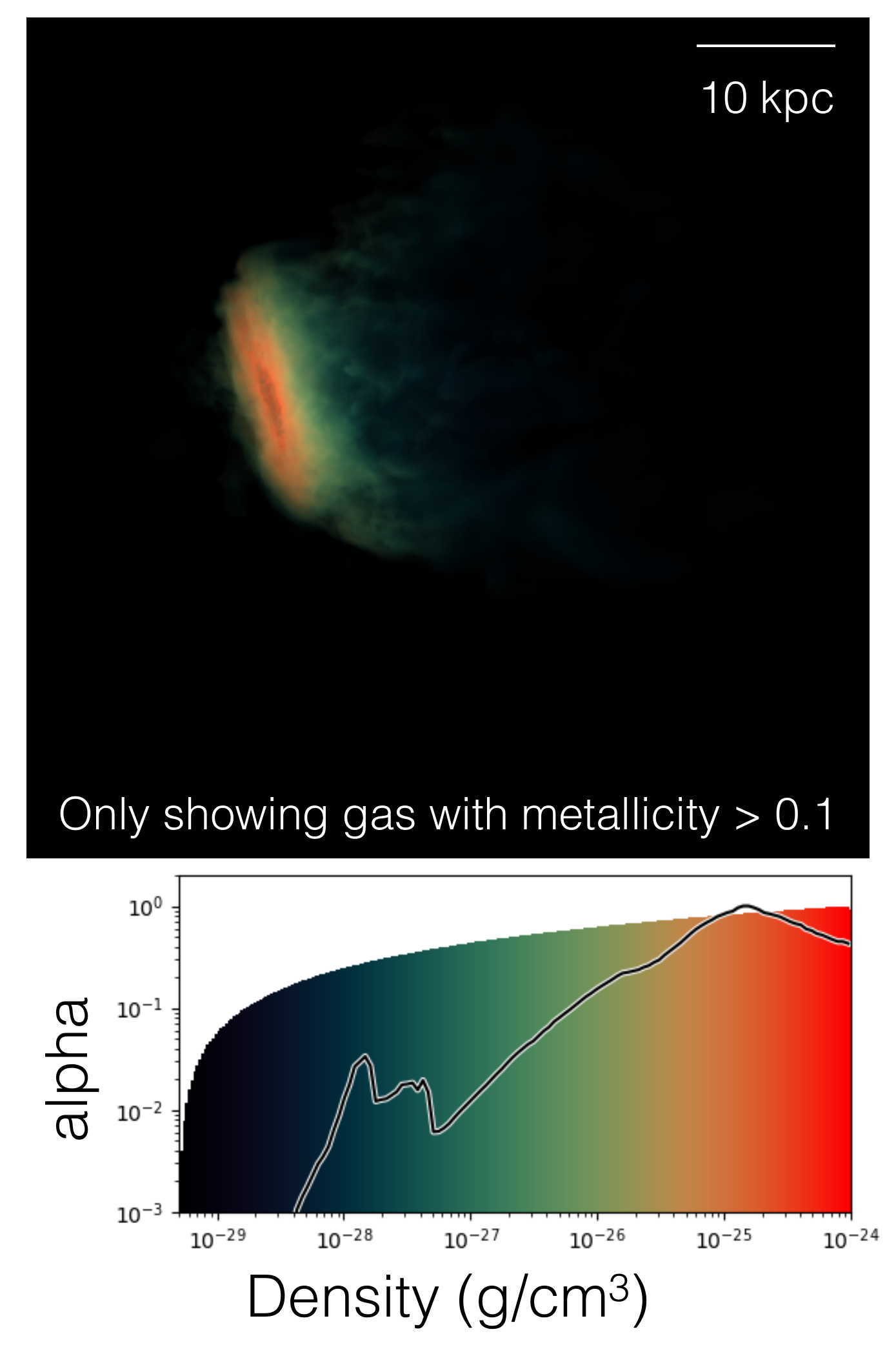}
    \caption{3D density rendering of LMC outflows driven by advecting (left) and streaming (right) cosmic rays, with ram pressure included. The density histograms below show the mapping between color and density in these renderings, where alpha represents the opacity. This point-of-view shows the full extent of the outflow and ram pressure contributions to a trailing filament, which is only of order 10 kpc long due to shielding of outflow gas by the LMC's predominantly face-on infall inclination. The simulation with cosmic ray streaming retains much more of a disk structure, as less gas is expelled into the LMC halo.}
    \label{fig:densRendering}
\end{figure*}

Our results for cosmic ray advection and cosmic ray streaming simulations are both shown in Figure \ref{fig:densRendering}, showing a volume rendering of the ISM gas (with metallicity greater than 0.1). The viewpoint is a 90 degree rotation compared to the line-of-sight, which is meant to easily show the stripped gas behind the LMC, which is in-falling mostly face-on until the most recent few hundred Myrs. We see a pronounced bow-shape formed in the disk, with the outflows propagating to the left (towards the inflow) being blown back towards the disk. To the right of the disk, the expelled gas flows unimpeded and mostly sheltered from ram pressure by the LMC disk itself. Only in the last few hundred Myrs, when the LMC tilts towards edge-on relative to the headwind, does the outflow gas significantly interact with ram pressure. As this is not enough time for ram pressure to push the outflow gas column significantly downstream, the tail that forms is only of order 10 kpc long. Viewed along the line-of-sight and decomposed into neutral and ionized hydrogen (Figure \ref{MockObservationFigure}), this filament is not easily visible, with only a small amount of additional gas in the upper right corner where the ram pressure stripped gas also resides. 

\begin{figure*}
    \centering
    \includegraphics[width = 0.8\textwidth]{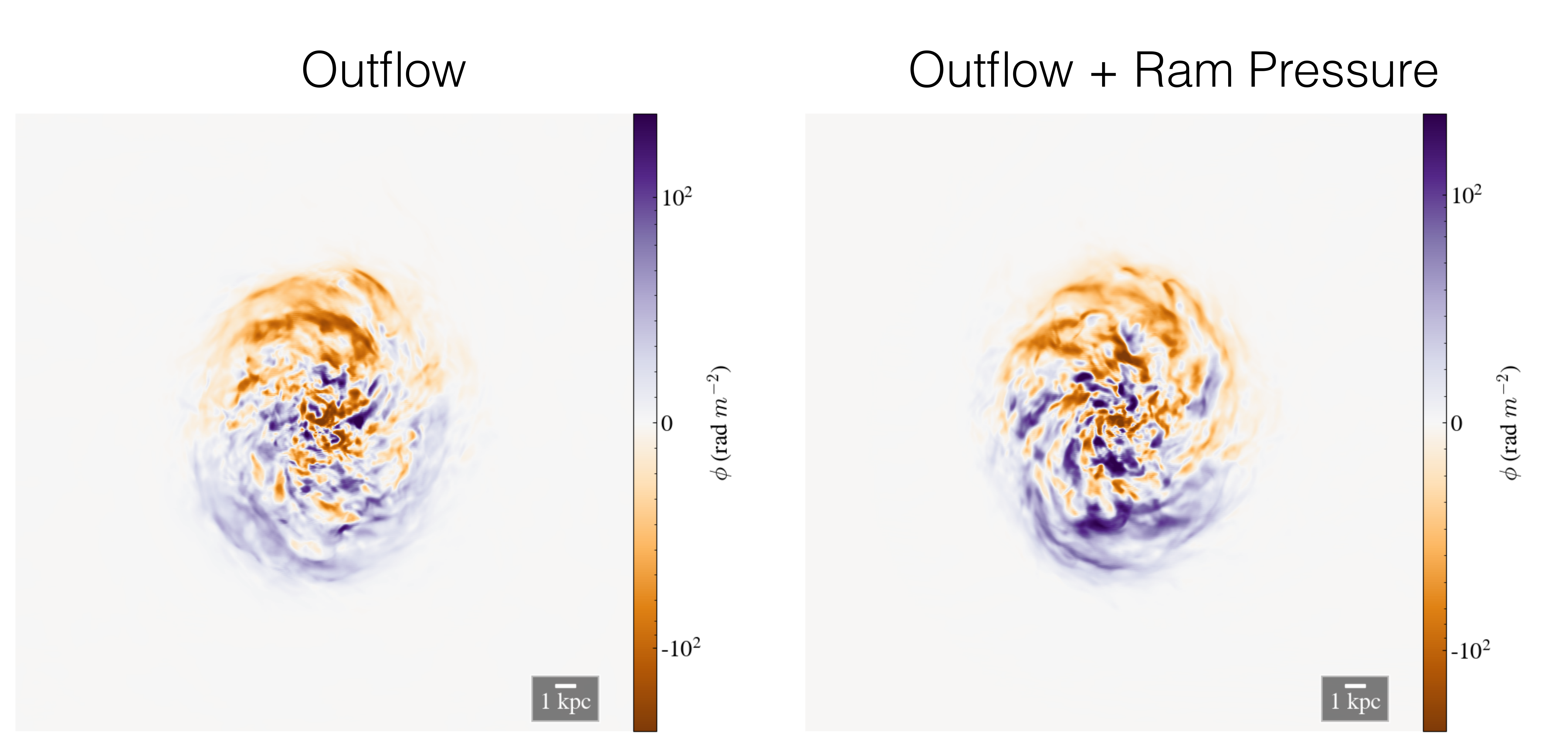}
    \caption{Present-day Faraday rotation measure of our simulated LMC outflow driven by cosmic ray advection, with and without ram pressure. No noticeable signature of a magnetized filament appears when ram pressure is included, even when we change the colorbar to focus on rotation measures between 1 and 10 (not shown). The rotation measure amplitude is maybe a bit higher near the leading edge (bottom left) when ram pressure is included, which would likely be due to compression. Overall, the two maps look quite similar.}
    \label{fig:Faraday1}
\end{figure*}

Figure \ref{fig:Faraday1} shows synthetic Faraday rotation measure maps for a present-day LMC after strong cosmic ray (advection) driven outflows, with and without ram pressure. To make this figure, we project along the present-day LOS to calculate the following quantity:

\begin{equation}
    \phi = (0.812 \rm rad/m^{2}) \int \frac{n_{e}(\rho, T)}{1 \rm cm^{-3}} \frac{B_{{\bf l}}}{1 \mu G} \frac{d{\bf l}}{1 \rm pc}
\end{equation}
where $B_{{\bf l}}$ is the magnetic field strength along the current LOS and the electron number density $n_{e}(\rho, T)$ is tabulated by \cite{Wiersma2009ThePlasmas} assuming photoionization equilibrium with the extragalactic UV background. The RM varies from roughly -250 to + 250 $\rm rad/m^{2}$, which if divided by a factor of 4-5 would match well with the spread measured by \cite{2012ApJ...759...25M}. There is no clear sign of a trailing, magnetized filament though, even when we vary the colorbar to focus on rotation measures less than 1-10. It looks like the leading edge (lower left) of the LMC might have a larger rotation measure amplitude when ram pressure is included, which would be consistent with compression amplifying the field. 

\begin{figure}
    \centering
    \includegraphics[width = 0.45\textwidth]{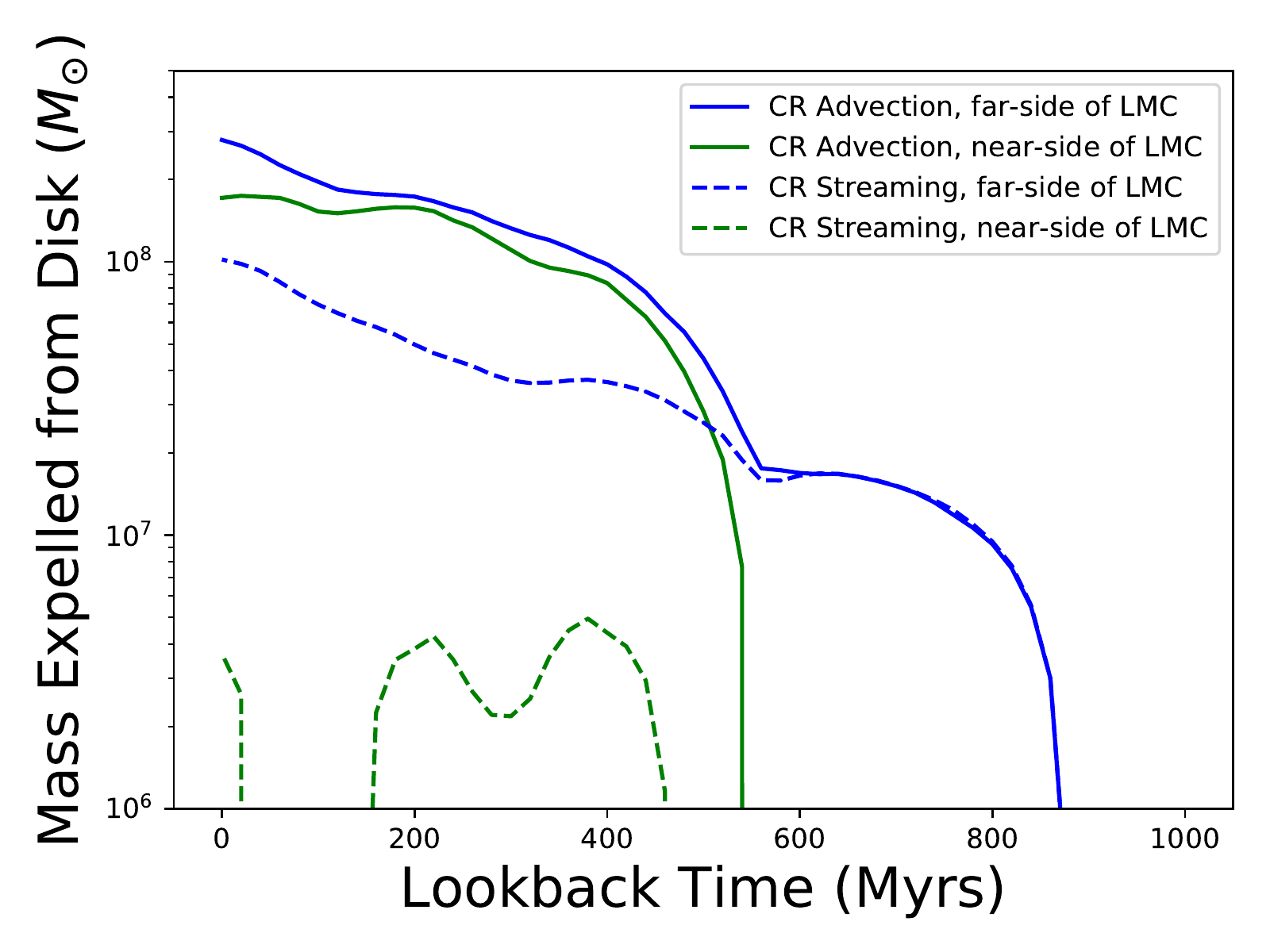}
    \caption{Mass expelled from both far and near sides of the LMC disk (positive or negative z-coordinate) for our cosmic ray driven wind plus ram pressure simulations. We see a large discrepancy, as ram pressure directly pushes against outflow escape, leading to far more gas in the LMC halo on the far-side of the disk. This suggests that absorption line studies probing only the foreground LMC halo could be severely underestimating the LMC halo mass, if it is formed by outflows.}
    \label{fig:near_far}
\end{figure}

More quantitatively, we see in Table \ref{tab:table} that, while the amount of gas outside the disk is significant, again not much gas is expelled from the larger LMC halo region, with less than $10^{7} M_{\odot}$ of additional gas expelled beyond a 13 kpc radius compared to the purely ram pressure simulation. In fact, the total amount of gas now expelled from the disk is even a bit lower than when outflows proceeded without ram pressure. This was seen in B2018 as well, as the increasingly heavy Milky Way halo gas flowing over the LMC disk at late times suppressed the outflow's vertical extent. 

In these new simulations with a 3D ram pressure, some of the ram pressure is working in direct opposition to outflows trying to break out of the LMC's near-side. In Figure \ref{fig:near_far}, we show the gas mass expelled from both the near and far sides of the disk as a function of time. Because the near-side wind encounters a direct hit from ram pressure, we would expect that less mass is expelled compared to the LMC's far-side, and this is confirmed. The effect can be quite dramatic, with the cosmic ray advection outflow lofting 3 times more gas on the far-side, while the streaming outflow only barely punches into the near-side halo when a burst of star formation occurs. This has intriguing consequences for interpreting absorption line studies of the LMC halo (e.g. \cite{Howk2002, Lehner2007, Barger2016}). \emph{Studies only probing the foreground may be severely underestimating the mass deposited in the near-side LMC halo by outflows.} Conversely, if a significant gas reservoir \emph{is} detected on the near-side, it suggests either a prevalence of recent outflows when the LMC is more favorably tilted edge-on, that past outflows have been strong enough to fight through the on-coming ram pressure, or that a pre-existing LMC halo still persists despite the effects of ram pressure and tidal stripping.  

We conclude from this that disk inclination is very important, with the recent face-on infall of the LMC being conducive to gas shielding instead of gas stripping. This, of course, does not mean that the lofted outflow gas cannot be stripped instead by tidal stripping from the SMC, which is not modeled here. We also only consider outflows from the most recent billion years. It is possible that more distant outflows, triggered by enhanced star formation episodes during, for instance, past interactions between the LMC and SMC, could expel a significant amount of gas into the Trailing Stream. Given the significant tidal effects of the SMC, this scenario is best-modeled with a full LMC-SMC interaction simulation, ideally including cosmic ray feedback and gas cooling, as we've done in these LMC-only simulations. 

\begin{figure*}
    \centering
    \includegraphics[width = 0.3\textwidth]{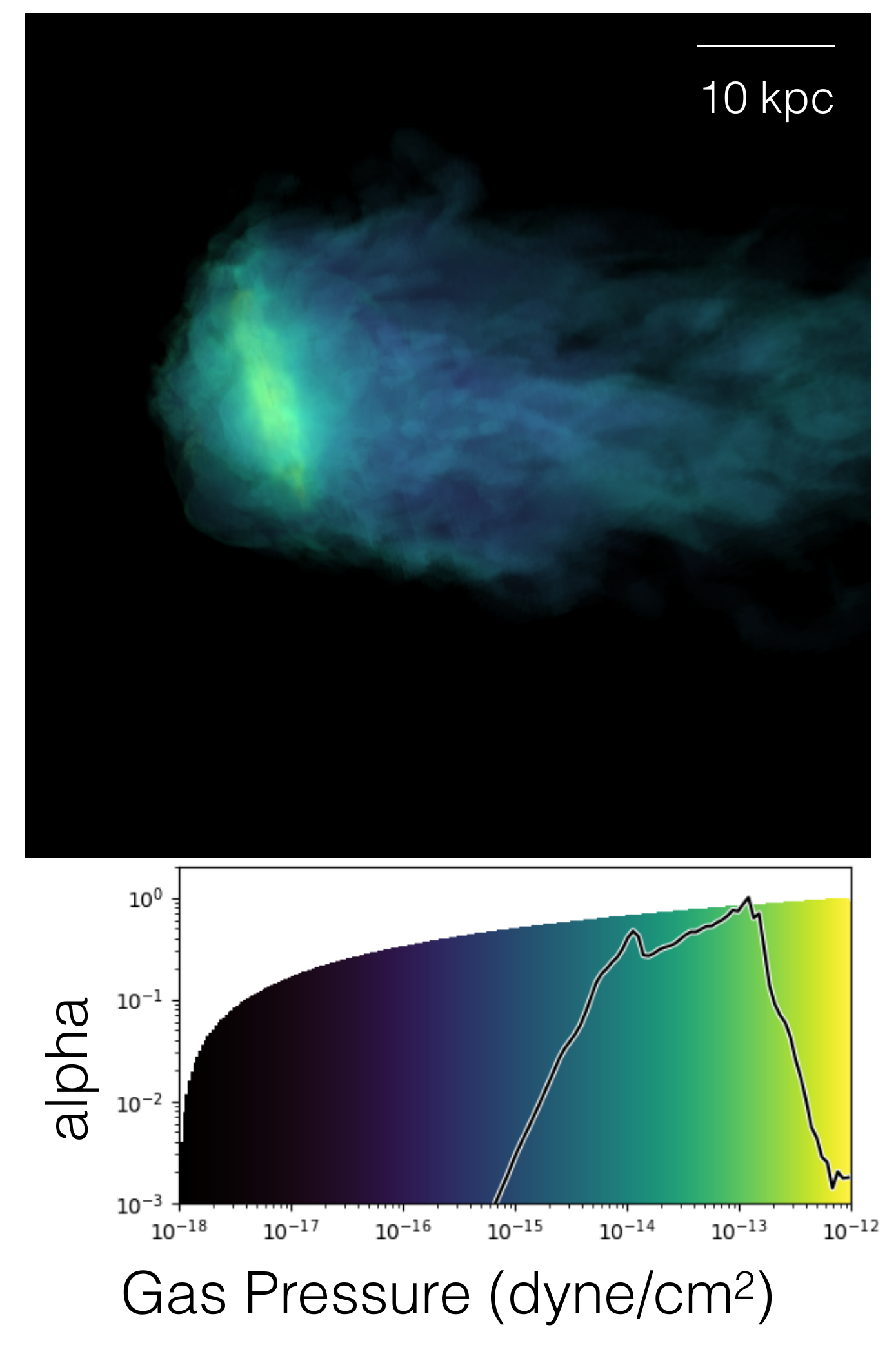}
    \includegraphics[width = 0.3\textwidth]{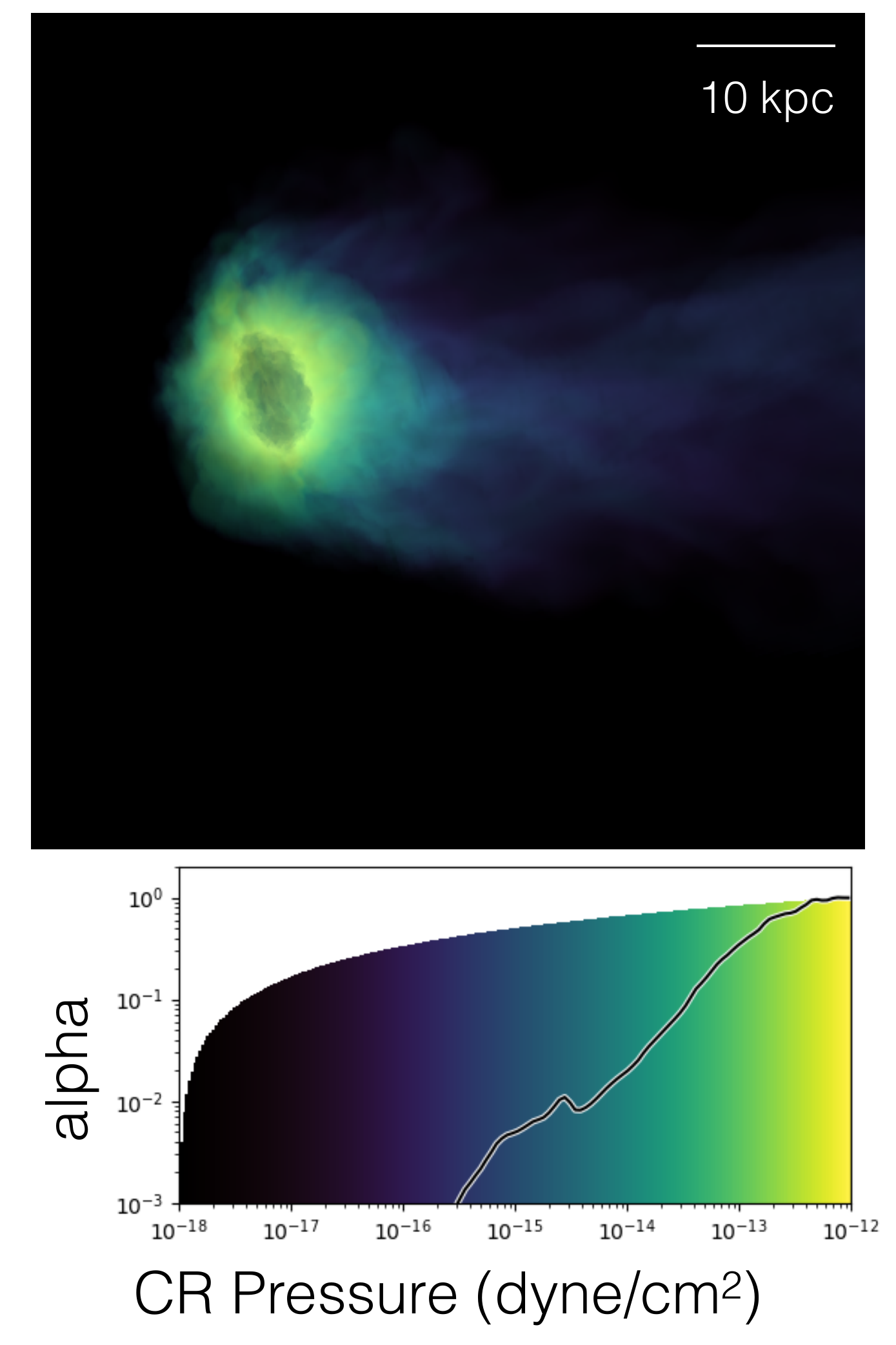}
    \includegraphics[width = 0.3\textwidth]{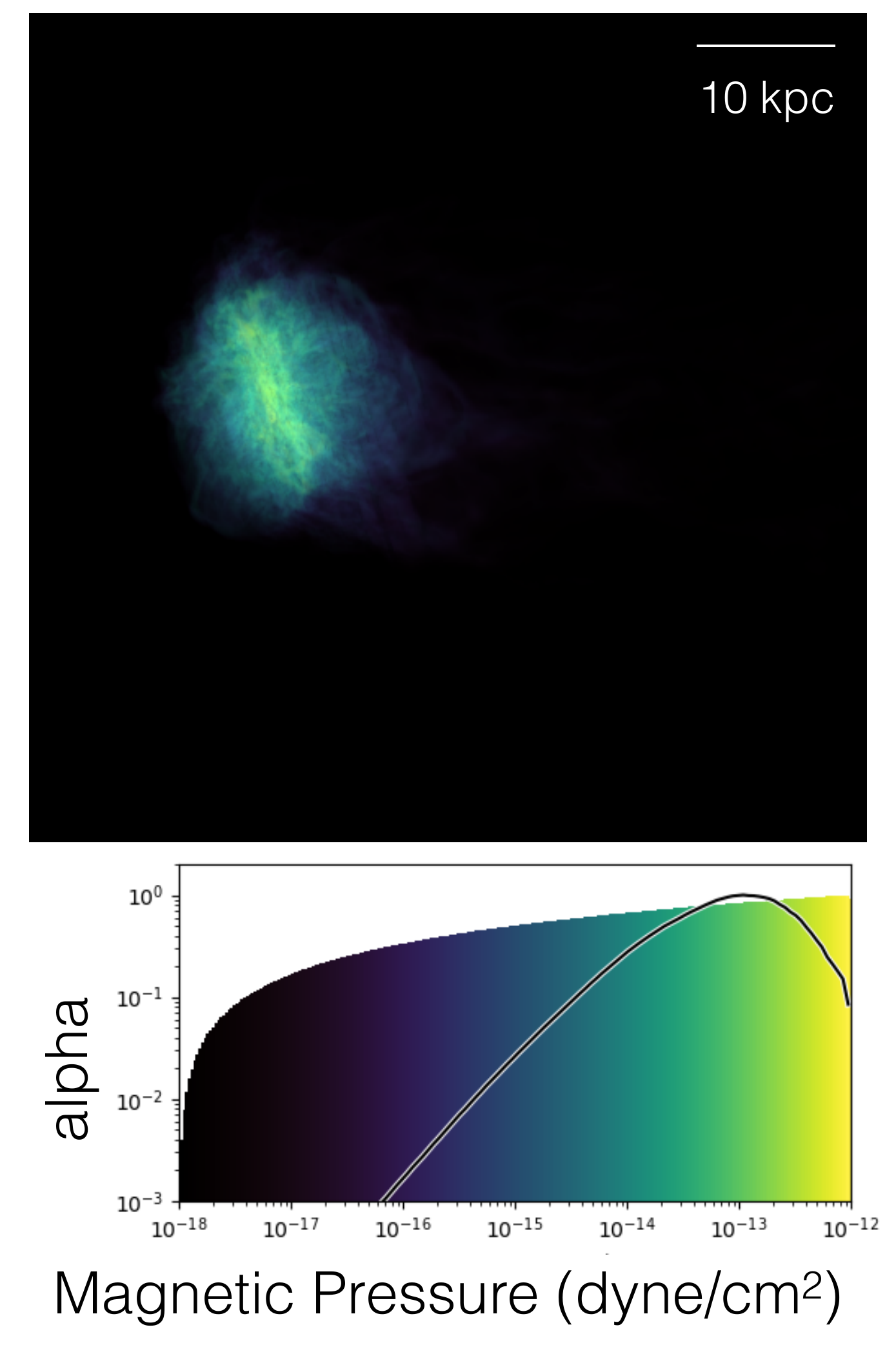}
    \includegraphics[width = 0.3\textwidth]{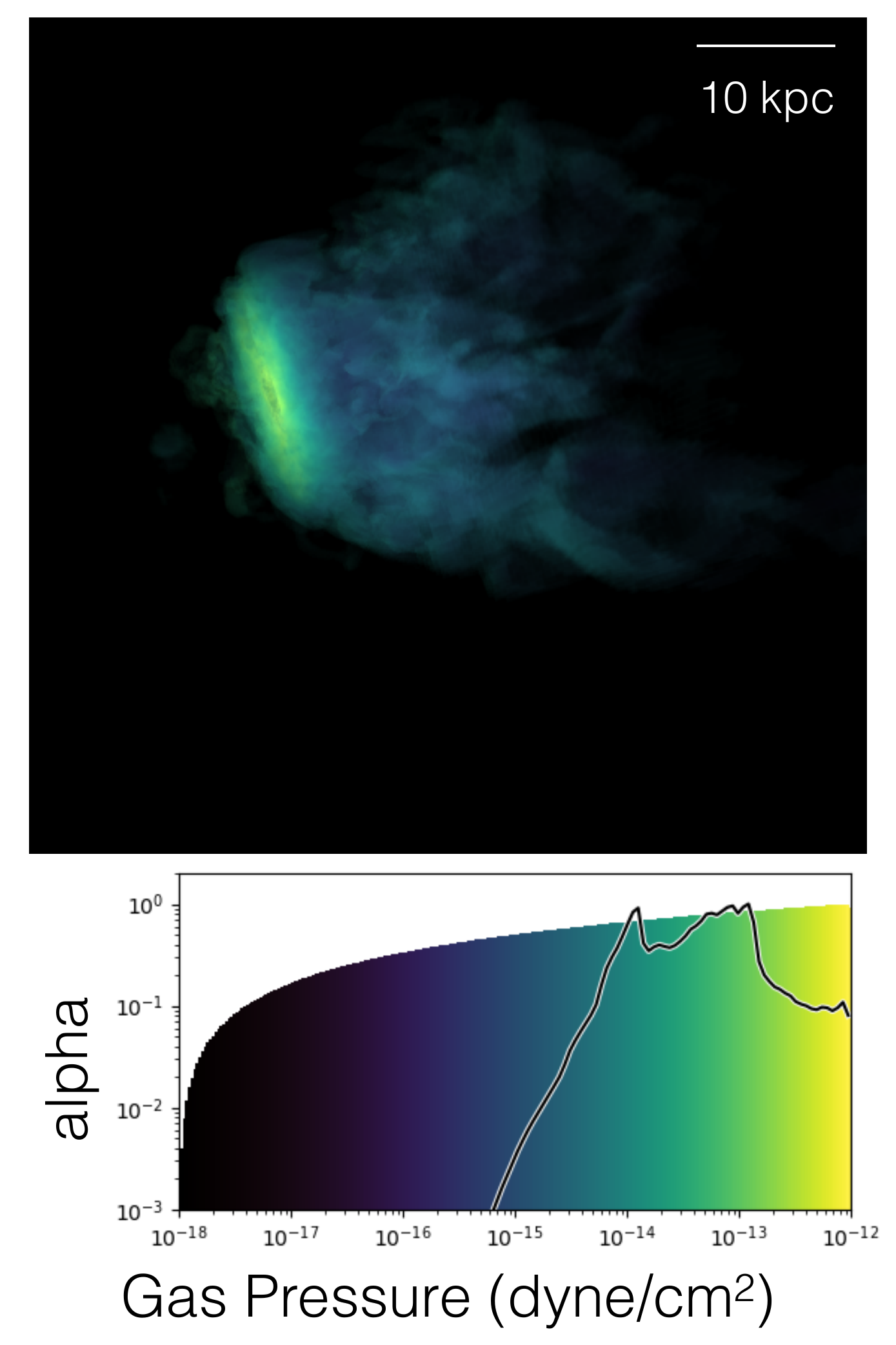}
    \includegraphics[width = 0.3\textwidth]{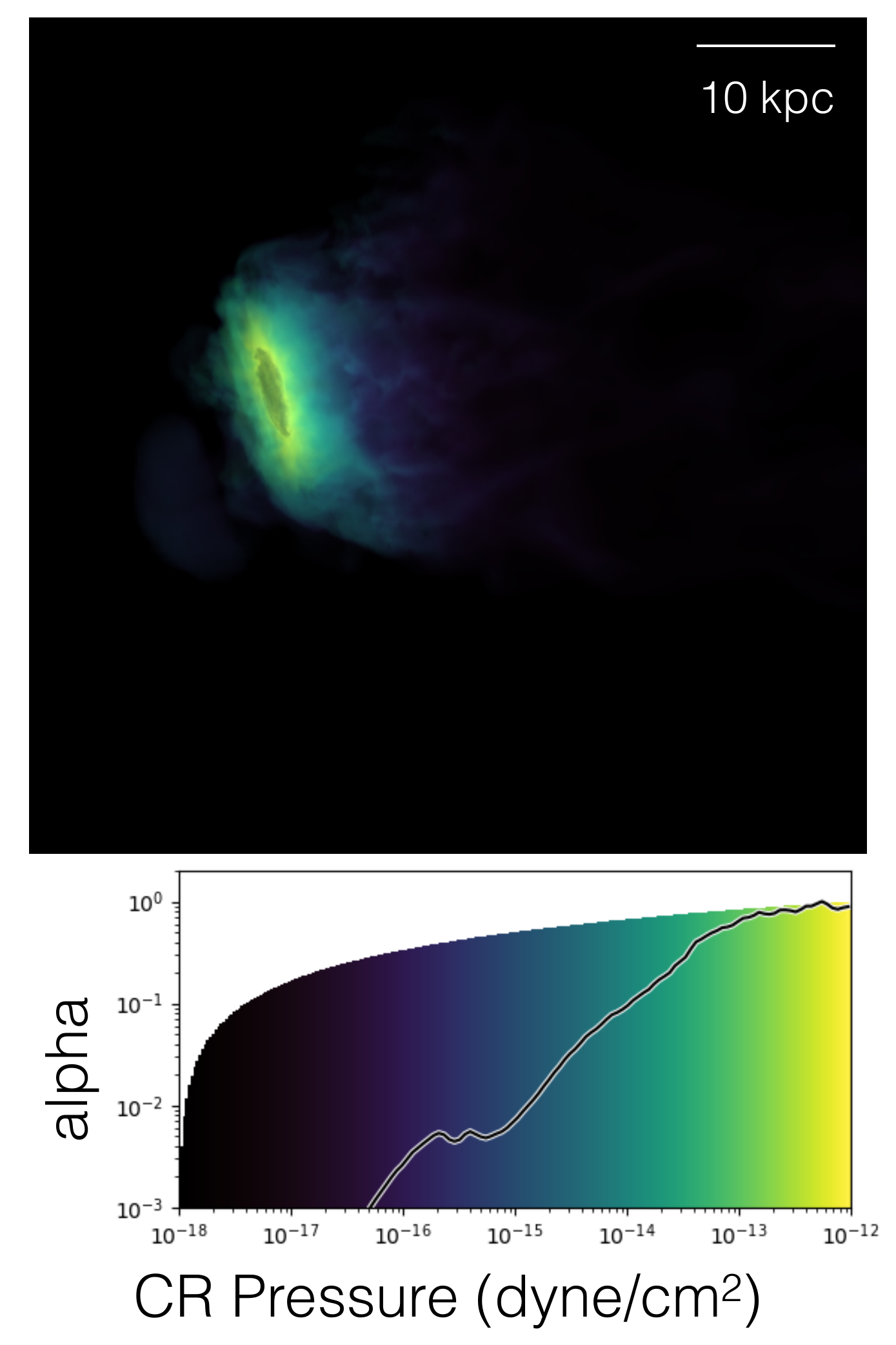}
    \includegraphics[width = 0.3\textwidth]{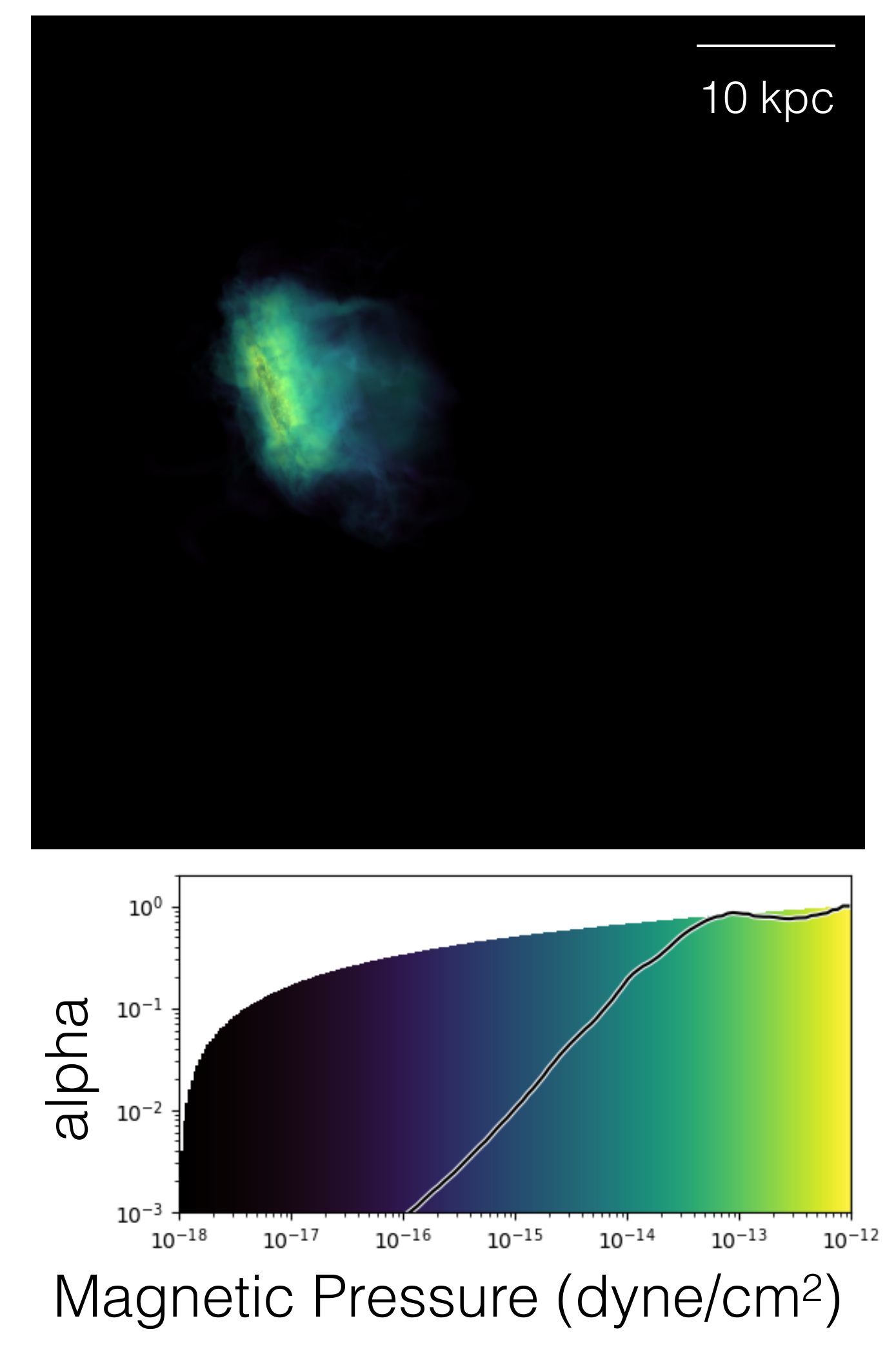}
    \caption{3D volume renderings of gas pressure (left), cosmic ray (CR) pressure (middle), and magnetic pressure (right), with normalized histograms below showing the mapping between pressure, color, and opacity (alpha). The top figures are for the cosmic ray advection case, while the bottom figures show the cosmic ray streaming plus losses case. Ram pressure is included in both simulations. Cosmic ray pressure clearly dominates gas and magnetic pressure in each case, especially the advection simulation where cosmic rays do not transfer energy to magnetic waves or lose energy in hadronic or Coulomb collisions. The distinct tail structure now noticeable in the gas pressure figures is primarily due to mixing of cold ISM and hot Milky Way halo gas, which increases the ISM tracer fraction (we only show gas with a fraction $>$ 0.01 here). Cosmic ray pressure begins to form a tail as well, which may be more pronounced in disks of different inclinations. }
    \label{fig:presRendering}
\end{figure*}

We note that outflows also affect the composition of the halo gas that will be stripped. In addition to expelling metals into the galaxy halo, and hence the ram pressure tail, outflows also project significant magnetic field and cosmic ray contributions above the disk. Figure \ref{fig:presRendering} shows 3D volume renderings of gas, cosmic ray, and magnetic pressure for our cosmic ray outflow plus ram pressure stripping simulations at present-day. We see that magnetic pressure is comparable to gas pressure within the disk region, and cosmic ray pressure greatly exceeds gas pressure throughout the whole galaxy region, especially in the halo, regardless of advection or streaming transport. The effects of such a cosmic ray dominated halo on ram pressure stripping has not, to our knowledge, been studied. Because cosmic rays constitute a relativistic fluid, the compressibility of the hybrid thermal gas - cosmic ray medium will change, likely affecting the formation and evolution of the ram pressure tail. Future simulations of galaxies at different inclination angles, or possibly of the LMC-SMC system, for which a tail \emph{does} form, will be able to address this. 




\subsection{Model Limitations}
As with any simulation of feedback and galaxy evolution, which attempt to connect vastly multi-scale outflow generation to observed galaxy properties, our simulations have assumptions and limitations. Here we expound upon a few that we think are most important. 

Our feedback implementation (see the Appendix) uses fitting functions of thermal and kinetic energy deposition motivated by simulations of isolated supernovae in inhomogeneous media \citep{Martizzi2015SupernovaMedium}. As each of our active particles represents a cluster, many times with mass greater than 1000 $M_{\odot}$, this energy is scaled up by the number of type II supernovae appropriate for that cluster. Clustering of supernovae, though, does not result in such a simple scaling: in fact, we might be underestimating the overall momentum injected into surrounding cells, as overlapping supernovae can boost the momentum by a factor of 4 or greater \citep{Gentry2017}. We find for our no cosmic ray simulations, however, that a factor of 5 momentum boost per cluster particle still cannot drive gas out of the disk.  This doesn't mean that thermally driven outflows cannot be driven in reality from the LMC, but this does strengthen one of our main results: the relative success of cosmic ray feedback.

This underestimation of thermally and kinetically driven outflows is likely compounded by resolution effects, as well, which mix the outflow hot phase too efficiently with warm ambient gas, thereby increasing the suppression of outflows by radiative cooling. This also affects the phase balance of outflowing gas. Outflow gas is susceptible to photoionization from at least the meta-galactic UV background, which is included in our simulations, but the detailed phase balance is also determined by the mass and energy load of the wind. Determining these parameters is an active topic of research, which requires an exploration of ISM feedback processes unresolved in these and most other galaxy-scale simulations.  
We also see future possibilities for improvement by including a pre-existing LMC gaseous halo with mass of order a few $\times 10^{9} M_{\odot}$, motivated by recent evidence for a high-mass dark matter halo $> 10^{11} M_{\odot}$ \citep{penarrubia2016, erkal2018, erkal2019}.  The extent to which this halo would suppress outflows or instead be punctured or entrained by them is unclear. Future observations are needed to estimate the LMC halo density in comparison to the Milky Way halo density. We briefly discuss implications for this LMC halo in Section \ref{conclusions}.

\section{Discussion and Conclusions}
\label{conclusions}

In this paper, we built upon previous work (B2018) to simulate outflows from the LMC, seeded by the derived star formation history of the LMC \citep{Harris2009THECLOUD} and energized by thermal, kinetic, and cosmic ray feedback. This serves a dual purpose to not only characterize the role of outflows in the Magellanic System but also to teach us about gas flows, cosmic ray transport, and feedback more generally. As a first application of our model, we revisited whether recent outflows could contribute to the LMC filament in the Trailing Magellanic Stream. We did this by simplifying the Magellanic System down to two components: outflows from the LMC and ram pressure stripping due to the LMC infall into the Milky Way halo. We modeled ram pressure as a time-varying headwind, changing direction over the last Gyr, in the frame of the LMC following the inclination, density, and velocity used by \cite{Salem2015RAMMEDIUM}. We simulated outflows from a low gas mass LMC and for a high gas mass LMC without cosmic rays included, with loss-less, advecting cosmic rays locked to the thermal gas, and also with additional cosmic ray streaming and collisional loss terms. Especially for the low gas mass disk with outflows driven by loss-less cosmic rays, since we only formed star particles when the predetermined star location has a gas density exceeding $10^{-25} g/cm^{3}$, the simulation star formation rate was far below the observed and drives an outflow that is unrealistically strong given the present-day mass flux and velocity estimates from \cite{Barger2016}. 

When we increased the gas mass of the LMC to account for not just neutral hydrogen but also the large reservoir of ionized gas, and when we included the more realistic cosmic ray treatment with streaming and collisional energy losses, the star formation better matched the derived rate and generated a more reasonable outflow. Most of the mass flux through the disk-halo interface occurs 400-600 Myrs in the past, with another large outburst occurring within the last 200 Myrs, generally following the star formation rate trends. Without ram pressure, the total gas displaced from the disk (defined as a disk of radius 13 kpc and height 1.7 kpc) is $7.80 \times 10^{7} M_{\odot}$. Future work is needed, however, to further constrain these outflows, especially using comparisons to gamma-ray observations, which gives us a sense of cosmic ray calorimetry in the LMC and a handle on the appropriate cosmic ray advection and streaming speeds out of the disk. 

Our main result is that, even for strong outflows that unbind $5 \times 10^{8} M_{\odot}$ from the LMC disk, gas is not easily swept away into the Trailing Stream. Because of the LMC's mainly face-on infall until the last few hundred Myrs, coinciding with a large boost in star formation that drives outflows between 400 and 600 Myrs ago, most of the outflow gas is shielded from ram pressure by the LMC disk. Mock neutral and ionized hydrogen column density maps along the line-of-sight show only small differences, compared to the same maps for the solely ram pressure simulation, in the gas contribution protruding from the LMC disk. Synthetic Faraday rotation measure maps similarly do not show a clear sign of trailing magnetized gas, even down to an amplitude of 1 rad $m^{-2}$.

This is not the end of the story, however. While this lofted, mostly ionized gas is trapped in the LMC halo in our models, tidal forces from the SMC, which we don't account for, may be able to more easily strip this gas now that feedback has projected it out of the gravitational potential well. These halos, fed partially by outflows that may have been prevalent for both the LMC \emph{and} SMC given strong outflow evidence for each galaxy \citep{Barger2016, McClure2018SMC} and past star formation bursts due to their interactions \citep{Harris2004SMC, Harris2009THECLOUD}, could represent a simple enhancement of features already created in tidal-only models. 


Our idealized simulations probe how much mass cosmic ray driven outflows could contribute to an LMC halo, both on the near and far sides of the disk, with the caveat that we do not include a pre-existing LMC halo that could suppress outflow breakout. We implore simulators of the full LMC-SMC tidal interaction to also include feedback, as the replenishment and pressurization of such a Magellanic halo, in the presence of tidal stripping, would be of great interest. To that end, the additional effects of the cosmic ray population, which significantly promote wind driving in our simulations, may be a crucial ingredient. 

While the interplay between ram pressure and outflows did not generate a large trailing filament in these simulations, another outflow-harboring galaxy at a more edge-on infall inclination may exhibit a significant mass expulsion that requires both ram pressure and outflows to expel the gas. In cosmological simulations, outflows from satellite galaxies represent a significant mode of gas transfer between the dwarf galaxy population and the host galaxy CGM and disk itself \citep{Angles2017, Hafen2019}. We argue that such outflows, aided by ram pressure, could also project an energetically significant cosmic ray population into the CGM.

Indeed, we found that the resulting LMC halo pressure in both our cosmic ray advection and streaming simulations was dominated by cosmic ray pressure. This naturally occurs because cosmic rays stream along vertical magnetic field lines away from the disk, where they can reside without significant energy losses in the diffuse halo. The creation of such cosmic ray dominated halos is supported by recent simulations \citep{Salem2016, Ji2019CRs}, especially of Milky Way mass galaxies, but as evidenced by this work, also possible for more massive dwarf galaxies. This cosmic ray presence can support more volume-filling cold gas than thermal pressure \citep{2018ApJ...868..108B, Ji2019CRs}, which primarily confines cold gas to clumps and filaments. This could leave imprints in absorption spectra, which needs to be studied further to disentangle cosmic ray driven vs thermally driven winds. If there were such a smoking gun indicator of cosmic ray driven winds vs thermally driven winds, the LMC may be a natural, nearby galaxy to test that theory with observations. 


\acknowledgements
The authors thank the anonymous referee for a thorough and helpful report that strengthened this paper. We also thank Greg Bryan, Andrew Fox, Naomi McClure-Griffiths, Ann Mao, Eve Ostriker, and Vadim Semenov for helpful discussions. We are also grateful to Mateusz Ruszkowski and Karen Yang for developing and providing us with their cosmic ray module in FLASH, which made this work possible. The software used in this work was in part developed by the DOE NNSA-ASC OASCR FLASH Center at the University of Chicago. This work made use of the \emph{yt} \citep{ytPaper}, Trident \citep{TridentRef}, and Astropy \citep{astropy} software packages, and we thank the developers for making those tools open-source. This work used the Extreme Science and Engineering Discovery Environment (XSEDE), which is supported by National Science Foundation grant number ACI-1548562 \citep{xsede}. Specifically, our computing resources stemmed from allocations TG-AST170033 and TG-AST190019 on the Stampede2 supercomputer. Simulations were also run on the University of Wisconsin - Madison HPC cluster. C.B. was partially supported by the National Science Foundation Graduate Research Fellowship Program under Grant No. DGE-1256259. E.D.O. acknowledges the Center for Computational Astrophysics at the Flatiron Institute for their hospitality during the completion of this work. We also acknowledge support from the Vilas Trust and the WARF Foundation at the University of Wisconsin, as well as NSF Grants AST-1616037 and PHY-1748958.

\appendix
\subsection{Frame Transformation}
The line-of-sight (LOS) frame is centered on the LMC's optical center of $(\rm RA, \rm DEC) = (82.24 \degree ,-69.5 \degree)$ \citep{vandermarel2002}. In this 3D Cartesian coordinate frame, the x-axis lies anti-parallel to the right ascension, the y-axis is parallel to the declination, and the z-axis is parallel to the line-of-sight to the observer, sitting in the solar neighborhood. Using the transformations defined in \cite{vandermarel2002}, we can switch to the ``LMC frame" using two rotations: a rotation by angle $\theta = 139.9 \degree$ about the LOS z-axis and a rotation by angle $i = 34.7 \degree$ about the new LMC frame x-axis. This puts us in a frame where the vertical axis is now aligned anti-parallel with the LMC's angular momentum axis, with the LMC mid-plane now in the LMC frame x-y plane. One further rotation about the new vertical axis by $100 \degree$ puts us in the ``simulation frame," where we no longer need to worry about the LMC headwind blowing into the box from more than 3 edges. Our simulations were all run in this frame, and mock observations were created by transforming back to the LOS frame, where our box can then be represented in RA-DEC coordinates by projecting into the Astropy \citep{astropy} World Coordinate System (WCS) centered on the LMC kinematic center. 

\subsection{Star Cluster Feedback}
In B2018, we probed the viability of outflows from the LMC by launching them in the least optimistic way: we injected thermal energy (neglecting cosmic rays and kinetic energy)  for a constant 30 Myr period of time into a ball of radius 100 pc, which is appropriate for a large star cluster such as 30 Doradus in the LMC. As is well-known from previous studies, the majority of this thermal energy is radiated away by line emission at temperatures near the peak of the cooling curve. This keeps the injected energy from building up a sufficient pressure gradient to blow gas out of the disk; however, with very clustered supernovae such as that assumed in B2018, a modest outflow or fountain may result. Our newly implemented method, based on the results of small patch simulations of supernovae in inhomogeneous media \citep{Martizzi2015SupernovaMedium}, includes both the thermal energy injection near the supernova and kinetic energy injection that still persists at large radii after the expanding supernova remnant shell has radiated away most of its thermal energy. This method is resolution-dependent and tunes the amount of thermal or kinetic energy injection into affected cells to give results consistent with the cooling radius determined from the \citet{Martizzi2015SupernovaMedium} simulations. To carry out this energy injection, we closely follow \cite{Semenov2017} and use active particles in FLASH to represent clusters of stars that evolve and explode over a 40 Myr time period, depositing energy and momentum to the particle cell and its surrounding cells according to the fitting functions of \citet{Martizzi2015SupernovaMedium} but scaled up by the number of type II supernovae expected for each cluster assuming a Chabrier IMF \citep{Chabrier2003}. Each individual supernova deposits the standard $10^{51}$ ergs of energy into surrounding cells, split between thermal, kinetic, and cosmic ray energy. When injecting cosmic rays, $10\% $ of the $10^{51}$ ergs are given to cosmic ray energy. The mass that each cluster particle ejects into the surrounding medium is determined following the prescription of \cite{Leitner2011}.


At high resolution, the Sedov-Taylor phase becomes more resolved, and this feedback prescription converges towards an entirely thermal energy deposition. In this case, the expansion to the snowplow phase is no longer sub-grid, and the momentum kick given to the surroundings is a direct outcome of the simulation. For high cluster masses, the temperature can exceed $10^{9}$ K when all energy deposition is thermal; therefore, for computational practicality, we impose a flag to artificially add mass to cells that will exceed $5 \times 10^{8}$ K. This artificial mass addition is done such that the total pressure in the cell will be consistent, but the sound speed (and hence, timestep) will be limited. Changing this temperature cutoff to $10^{9}$ K does not significantly change our results.

\subsection{Towards Higher Resolution}

\begin{figure*}
    \centering
    \includegraphics[width = 0.4\textwidth]{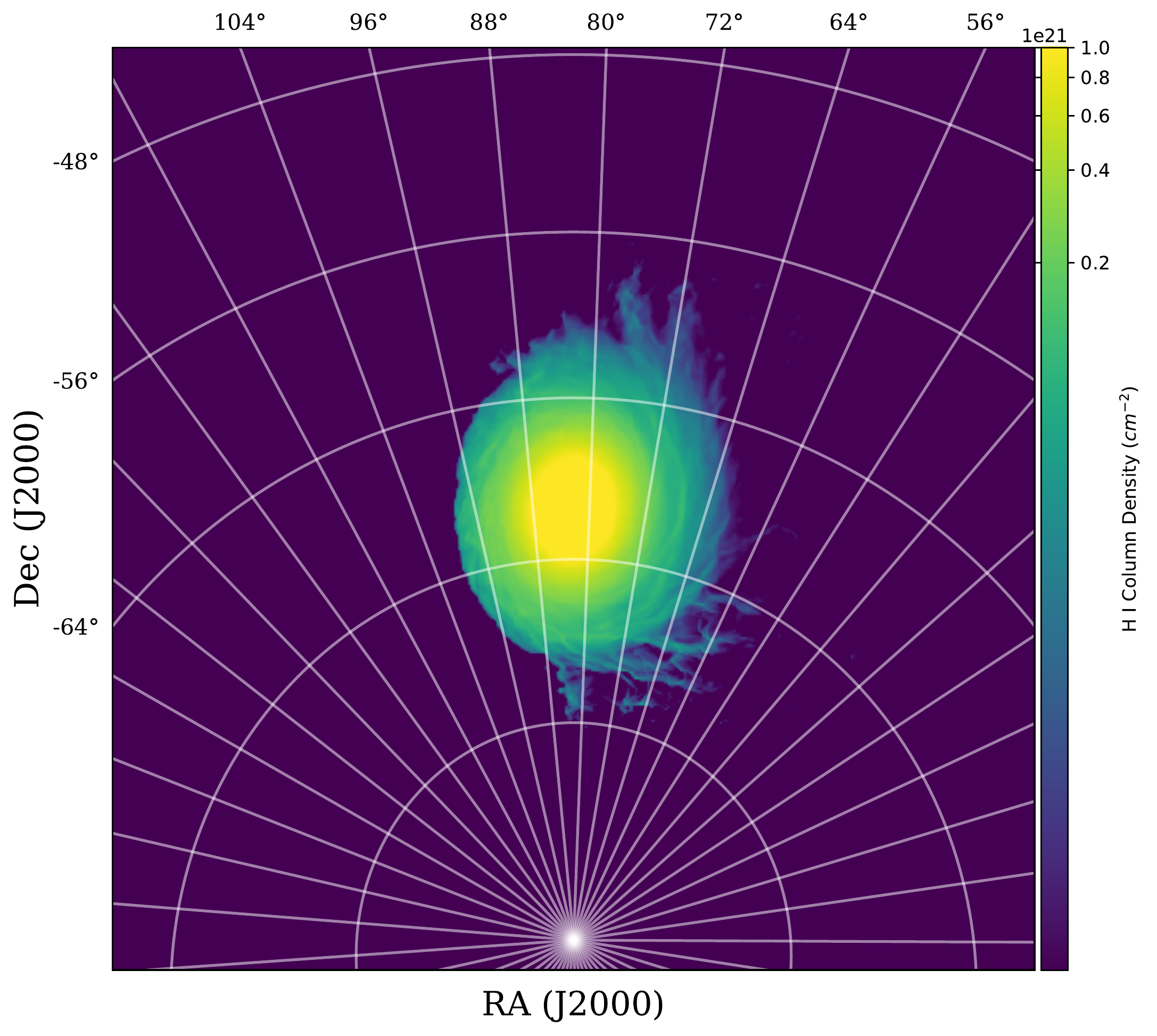}
    \includegraphics[width = 0.4\textwidth]{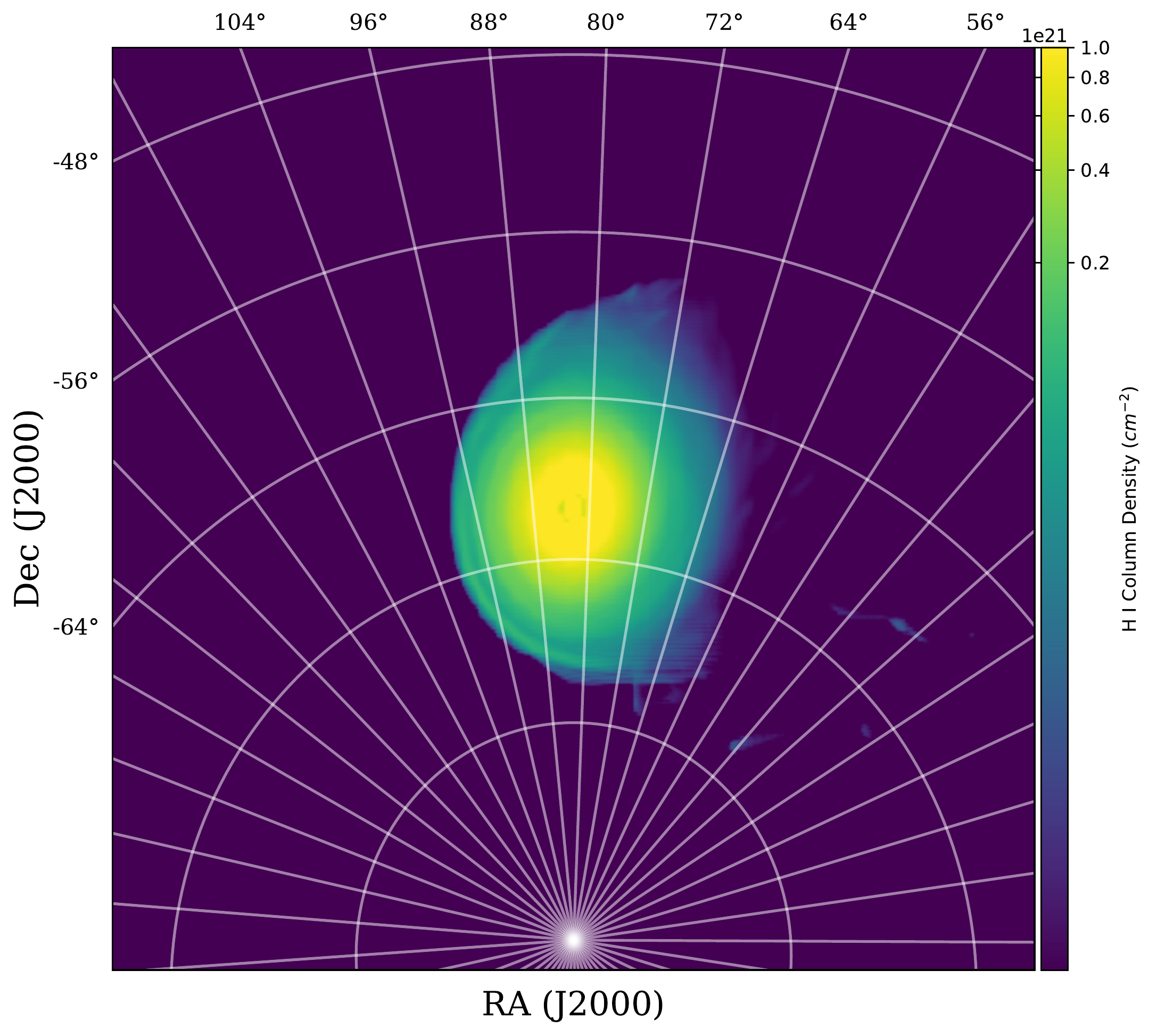}
    \caption{H I column densities after ram pressure stripping (no outflows) of the low gas mass LMC at maximum resolution of 39 pc (left) and 78 pc (right). More structure is clearly visible in the higher resolution run. }
    \label{fig:resolution1}
\end{figure*}

Here, we show part of our growing resolution study, with a combination of ram pressure only, outflow only, and outflow plus ram pressure simulations at maximum resolutions ranging from 39 to 156 pc. Because our code relies on numerical diffusivities, we consider this study as more of a probe of the effects of resolution rather than a convergence study. We generally find consistent results overall in terms of mass expulsion from the disk, which is a main driver of this work. For the cosmic ray (advection) driven outflow from the low gas mass LMC, $2.56 \times 10^{8} M_{\odot}$ is expelled when the maximum resolution is 78 pc, and $2.59 \times 10^{8} M_{\odot}$ is expelled at a resolution of 156 pc, representing only a 1\% change. At a maximum resolution of 78 pc, ram pressure stripping expels $1.36 \times 10^{7} M_{\odot}$ from the low gas mass LMC disk, while $1.63 \times 10^{7} M_{\odot}$ is expelled at a maximum resolution of 39 pc. This 20\% increase does not change our conclusion that ram pressure alone is inefficient, but the morphology of the stripped gas clearly changes with resolution. Higher resolution reduces mixing of cold clumps with the hot Milky Way halo (see Figure \ref{fig:resolution1}).

Similarly, our high gas mass LMC with ram pressure and cosmic ray driven outflows shows slightly more structure than its low resolution counterpart (not shown). Interestingly, the total mass expelled actually decreases at higher resolution ($5.26 \times 10^{8} M_{\odot}$ at 156 pc resolution and $4.50 \times 10^{8} M_{\odot}$ at 78 pc resolution), representing a decrease of 14\%. We are currently running this simulation at 39 pc and 20 pc resolution, as well, as future work comparing mock observables to the LMC relies on a better encapsulation of the mass and energy loading of outflows, which necessitates higher resolution. 

We also checked, at 156 pc resolution, how our preliminary streaming simulations changed as we varied the free parameters of our streaming implementation. Because, in the streaming picture, the flow along field lines is always directed down the cosmic ray pressure gradient, numerical issues arise near extrema in cosmic ray pressure, where the gradient changes sign. To counteract this we use a regularization method \citep{2009arXiv0909.5426S}, in which one chooses an appropriate scale length, L, for the system that then defines a characteristic cosmic ray pressure gradient, $P_{CR} / L$. The streaming speed of cosmic rays, which ideally is $v_{A}$, is approximated as 

\begin{equation}
    v_{s} = v_{A} \rm tanh \left( \frac{ | \mathbf{\hat{b}} \cdot  \nabla P_{CR} |}{P_{CR}/L} \right)
\end{equation}
For a cosmic ray pressure gradient $\nabla P_{CR}$ much greater than $P_{CR}/L$, $v_{s} = v_{A}$, while smaller cosmic ray pressure gradients will lead to $v_{s} < v_{A}$. One would like L to be as large as possible (so $v_{s} \approx v_{A}$ for a wide range of cosmic ray pressure gradients in the system); however, the proper simulation timestep constraint for this regularization method is $dt  < dx^{2}/(v_{s} L)$, which is second order in cell width. 

We tested scale lengths of L = 1, 5, and 10 kpc, and we found that the L = 5 and L = 10 kpc simulations were sufficiently converged in terms of mass expulsion and direct comparisons between cosmic ray pressure slices and projections. We choose to use L = 5 kpc, and for this value, the streaming timestep is almost always the limiting timestep. Future work to higher resolution below 40 pc may indeed require a method that scales better, such as a two-moment method \citep{2018ApJ...854....5J}.

We point out that cosmic ray evolution is also resolution-dependent due to numerical diffusivities. There are two separate issues: even if the cosmic ray population, under anisotropic diffusion or streaming, follows field lines exactly, numerical resistivity leads to errors in the \emph{magnetic field} evolution, and hence the cosmic ray evolution. Additionally, numerical algorithms for cosmic ray transport are not infinitely precise. Even purely advecting cosmic rays, which have no sense of the magnetic field geometry, are susceptible to resolution-dependent numerical diffusion. While it is difficult to estimate numerical diffusion, one should be aware of these effects when interpreting results.


\centering

\bibliographystyle{apj}
\bibliography{bibliography}
\end{document}